





         \font\twelvei=cmmi10 scaled 1200
  \font\twelvesy=cmsy10 scaled 1200

  \font\twelvemib=cmmib10 scaled 1200
  \font\elevenmib=cmmib10 scaled 1095
  
  \font\eightmib=cmmib10 scaled 800
  \font\sixmib=cmmib10 scaled 667


\font\elevenrm=cmr10 scaled 1095    \font\eleveni=cmmi10 scaled 1095
\font\elevensy=cmsy10 scaled 1095   \font\elevenex=cmex10 scaled 1095
\font\elevenbf=cmbx10 scaled 1095   \font\elevensl=cmsl10 scaled 1095
\font\eleventt=cmtt10 scaled 1095   \font\elevenit=cmti10 scaled 1095


\font\seventeeni=cmmi10 scaled \magstep3

\font\seventeensy=cmsy10 scaled \magstep3

\font\seventeenmib=cmmib10 scaled \magstep3

\newfam\cpfam%


\font\eightrm=cmr8 \font\eighti=cmmi8
\font\eightsy=cmsy8 \font\eightbf=cmbx8


\font\sixrm=cmr6 \font\sixi=cmmi6
\font\sixsy=cmsy6 \font\sixbf=cmbx6

\skewchar\eleveni='177   \skewchar\elevensy='60
\skewchar\elevenmib='177  \skewchar\seventeensy='60
\skewchar\seventeenmib='177
\skewchar\seventeeni='177

\newfam\mibfam%


\def\elevenpoint{\normalbaselineskip=12.2pt
  \abovedisplayskip 12.2pt plus 3pt minus 9pt
  \belowdisplayskip 12.2pt plus 3pt minus 9pt
  \abovedisplayshortskip 0pt plus 3pt
  \belowdisplayshortskip 7.1pt plus 3pt minus 4pt
  \smallskipamount=3.3pt plus1.1pt minus1.1pt
  \medskipamount=6.6pt plus2.2pt minus2.2pt
  \bigskipamount=13.3pt plus4.4pt minus4.4pt
  \def\rm{\fam0\elevenrm}          \def\it{\fam\itfam\elevenit}%
  \def\sl{\fam\slfam\elevensl}     \def\bf{\fam\bffam\elevenbf}%
  \def\mit{\fam 1}                 \def\cal{\fam 2}%
  \def\tt{\eleventt}
  \def\mib{\fam\mibfam\elevenmib}%
  \textfont0=\elevenrm   \scriptfont0=\eightrm   \scriptscriptfont0=\sixrm
  \textfont1=\eleveni    \scriptfont1=\eighti    \scriptscriptfont1=\sixi
  \textfont2=\elevensy   \scriptfont2=\eightsy   \scriptscriptfont2=\sixsy
  \textfont3=\elevenex   \scriptfont3=\elevenex  \scriptscriptfont3=\elevenex
  \textfont\itfam=\elevenit
  \textfont\slfam=\elevensl
  \textfont\bffam=\elevenbf \scriptfont\bffam=\eightbf
  \scriptscriptfont\bffam=\sixbf
  \textfont\mibfam=\elevenmib
  \scriptfont\mibfam=\eightmib
  \scriptscriptfont\mibfam=\sixmib
  \def\xrm{\textfont0=\elevenrm\scriptfont0=\eightrm
      \scriptscriptfont0=\sixrm}
  \normalbaselines\rm}

  \skewchar\twelvei='177   \skewchar\twelvesy='60
  \skewchar\twelvemib='177
%
%


\mathchardef\alpha="710B
\mathchardef\beta="710C
\mathchardef\gamma="710D
\mathchardef\delta="710E
\mathchardef\epsilon="710F
\mathchardef\zeta="7110
\mathchardef\eta="7111
\mathchardef\theta="7112
\mathchardef\kappa="7114
\mathchardef\lambda="7115
\mathchardef\mu="7116
\mathchardef\nu="7117
\mathchardef\xi="7118
\mathchardef\pi="7119
\mathchardef\rho="711A
\mathchardef\sigma="711B
\mathchardef\tau="711C
\mathchardef\phi="711E
\mathchardef\chi="711F
\mathchardef\psi="7120
\mathchardef\omega="7121
\mathchardef\varepsilon="7122
\mathchardef\vartheta="7123
\mathchardef\varrho="7125
\mathchardef\varphi="7127



\def\beginlinemode{\endmode
  \begingroup\parskip=0pt \obeylines\def\\{\par}\def\endmode{\par\endgroup}}
\def\beginparmode{\endmode
  \begingroup \def\endmode{\par\endgroup}}
\let\endmode=\par
{\obeylines\gdef\
{}}
\def\singlespace{\baselineskip=\normalbaselineskip}

\def\oneandahalfspace{\baselineskip=\normalbaselineskip
  \multiply\baselineskip by 3 \divide\baselineskip by 2}
\def\doublespace{\baselineskip=\normalbaselineskip \multiply\baselineskip by 2}

\nopagenumbers
\newcount\firstpageno
\firstpageno=2
\footline={\ifnum\pageno<\firstpageno{\hfil}\else{\hfil\elevenrm\folio\hfil}\fi}
\let\rawfootnote=\footnote		
\def\footnote#1#2{{\oneandahalfspace\parindent=0pt
\rawfootnote{#1}{#2}}}
\def\raggedcenter{\leftskip=4em plus 12em \rightskip=\leftskip
  \parindent=0pt \parfillskip=0pt \spaceskip=.3333em \xspaceskip=.5em
  \pretolerance=9999 \tolerance=9999
  \hyphenpenalty=9999 \exhyphenpenalty=9999 }
\def\dateline{\rightline{\ifcase\month\or
  January\or February\or March\or April\or May\or June\or
  July\or August\or September\or October\or November\or December\fi
  \space\number\year}}
\def\received{\vskip 3pt plus 0.2fill
 \centerline{\sl (Received\space\ifcase\month\or
  January\or February\or March\or April\or May\or June\or
  July\or August\or September\or October\or November\or December\fi
  \qquad, \number\year)}}


\hsize=6.5truein
\hoffset=0truein
\vsize=8.9truein
\voffset=0truein
\hfuzz=0.1pt
\vfuzz=0.1pt
\parskip=\medskipamount
\overfullrule=0pt	



\def\title			
  {\null\vskip 3pt plus 0.2fill
   \beginlinemode \doublespace \raggedcenter \bf}

\def\author			
  {\vskip 3pt plus 0.2fill \beginlinemode
   \singlespace \raggedcenter}

\def\affil			
  {\vskip 3pt plus 0.1fill \beginlinemode
   \oneandahalfspace \raggedcenter \sl}

\def\abstract			
  {\vskip 3pt plus 0.3fill \beginparmode
   \doublespace \narrower ABSTRACT: }

\def\summary			
  {\vskip 3pt plus 0.3fill \beginparmode
   \doublespace \narrower SUMMARY: }

\def\pacs#1
  {\vskip 3pt plus 0.2fill PACS: #1}

\def\endtitlepage		
  {\endpage			
   \body}

\def\body			
  {\beginparmode}		

\def\head#1{			
  \filbreak\vskip 0.5truein	
  {\immediate\write16{#1}
   \raggedcenter \uppercase{#1}\par}
   \nobreak\vskip 0.25truein\nobreak}

\def\refto#1{$^{#1}$}		

\def\references			
  {\head{References}		
   \beginparmode
   \frenchspacing \parindent=0pt \leftskip=1truecm
   \parskip=8pt plus 3pt \everypar{\hangindent=\parindent}}

\gdef\refis#1{\indent\hbox to 0pt{\hss[#1]~}}	

\gdef\journal#1, #2, #3, 1#4#5#6{		
    {\sl #1~}{\bf #2}, #3 (1#4#5#6)}		

\def\refstylenp{		
  \gdef\refto##1{ [##1]}				
  \gdef\refis##1{\indent\hbox to 0pt{\hss##1)~}}	
  \gdef\journal##1, ##2, ##3, ##4 {			
     {\sl ##1~}{\bf ##2~}(##3) ##4 }}

\def\refstyleprnp{		
  \gdef\refto##1{ [##1]}				
  \gdef\refis##1{\indent\hbox to 0pt{\hss##1)~}}	
  \gdef\journal##1, ##2, ##3, 1##4##5##6{		
    {\sl ##1~}{\bf ##2~}(1##4##5##6) ##3}}

\def\refstylejphys{		
  \gdef\refto##1{[##1]}				
  \gdef\refis##1{\indent\hbox to 0pt{\hss[##1]~}}	
  \gdef\journal##1, ##2, ##3, 1##4##5##6{		
     1##4##5##6 {\sl ##1~}{\bf ##2~}{##3}}}

\def\endreferences{\body}

\def\figurecaptions		
  {\endpage
   \beginparmode
   \head{Figure Captions}
}

\def\endpage			
  {\vfill\eject}

\def\endpaper			
  {\endmode\vfill\supereject}


\def\ref#1{Ref[#1]}			
\def\Ref#1{Ref[#1]}			
\def\Refs#1{Refs[#1]}			

\def\frac#1#2{{\textstyle{#1 \over #2}}}
\def\half{{\textstyle{ 1\over 2}}}

\def\sla{\raise.15ex\hbox{$/$}\kern-.57em}
\def\leaderfill{\leaders\hbox to 1em{\hss.\hss}\hfill}
\def\twiddle{\lower.9ex\rlap{$\kern-.1em\scriptstyle\sim$}}
\def\bigtwiddle{\lower1.ex\rlap{$\sim$}}
\def\gtwid{\mathrel{\raise.3ex\hbox{$>$\kern-.75em\lower1ex\hbox{$\sim$}}}}
\def\ltwid{\mathrel{\raise.3ex\hbox{$<$\kern-.75em\lower1ex\hbox{$\sim$}}}}
\def\square{\kern1pt\vbox{\hrule height 1.2pt\hbox{\vrule width 1.2pt\hskip 3pt
   \vbox{\vskip 6pt}\hskip 3pt\vrule width 0.6pt}\hrule height 0.6pt}\kern1pt}

\catcode`@=11
\newcount\r@fcount \r@fcount=0
\newcount\r@fcurr
\immediate\newwrite\reffile
\newif\ifr@ffile\r@ffilefalse
\def\w@rnwrite#1{\ifr@ffile\immediate\write\reffile{#1}\fi\message{#1}}

\def\writer@f#1>>{}
\def\referencefile{
  \r@ffiletrue\immediate\openout\reffile=\jobname.ref%
  \def\writer@f##1>>{\ifr@ffile\immediate\write\reffile%
    {\noexpand\refis{##1} = \csname r@fnum##1\endcsname = %
     \expandafter\expandafter\expandafter\strip@t\expandafter%
     \meaning\csname r@ftext\csname r@fnum##1\endcsname\endcsname}\fi}%
  \def\strip@t##1>>{}}

\def\citeall#1{\xdef#1##1{#1{\noexpand\cite{##1}}}}
\def\cite#1{\each@rg\citer@nge{#1}}	

\def\each@rg#1#2{{\let\thecsname=#1\expandafter\first@rg#2,\end,}}
\def\first@rg#1,{\thecsname{#1}\apply@rg}	
\def\apply@rg#1,{\ifx\end#1\let\next=\relax
\else,\thecsname{#1}\let\next=\apply@rg\fi\next}

\def\citer@nge#1{\citedor@nge#1-\end-}	
\def\citer@ngeat#1\end-{#1}
\def\citedor@nge#1-#2-{\ifx\end#2\r@featspace#1 
  \else\citel@@p{#1}{#2}\citer@ngeat\fi}	
\def\citel@@p#1#2{\ifnum#1>#2{\errmessage{Reference range #1-#2\space is bad.}
    \errhelp{If you cite a series of references by the notation M-N, then M and
    N must be integers, and N must be greater than or equal to M.}}\else%
 {\count0=#1\count1=#2\advance\count1
by1\relax\expandafter\r@fcite\the\count0,%
  \loop\advance\count0 by1\relax
    \ifnum\count0<\count1,\expandafter\r@fcite\the\count0,%
  \repeat}\fi}

\def\r@featspace#1#2 {\r@fcite#1#2,}	
\def\r@fcite#1,{\ifuncit@d{#1}		
    \expandafter\gdef\csname r@ftext\number\r@fcount\endcsname%
    {\message{Reference #1 to be supplied.}\writer@f#1>>#1 to be supplied.\par
     }\fi%
  \csname r@fnum#1\endcsname}

\def\ifuncit@d#1{\expandafter\ifx\csname r@fnum#1\endcsname\relax%
\global\advance\r@fcount by1%
\expandafter\xdef\csname r@fnum#1\endcsname{\number\r@fcount}}

\let\r@fis=\refis			
\def\refis#1#2#3\par{\ifuncit@d{#1}
    \w@rnwrite{Reference #1=\number\r@fcount\space is not cited up to now.}\fi%
  \expandafter\gdef\csname r@ftext\csname r@fnum#1\endcsname\endcsname%
  {\writer@f#1>>#2#3\par}}

\def\r@ferr{\endreferences\errmessage{I was expecting to see
\noexpand\endreferences before now;  I have inserted it here.}}
\let\r@ferences=\references
\def\references{\r@ferences\def\endmode{\r@ferr\par\endgroup}}

\let\endr@ferences=\endreferences
\def\endreferences{\r@fcurr=0
  {\loop\ifnum\r@fcurr<\r@fcount
    \advance\r@fcurr by 1\relax\expandafter\r@fis\expandafter{\number\r@fcurr}%
    \csname r@ftext\number\r@fcurr\endcsname%
  \repeat}\gdef\r@ferr{}\endr@ferences}


\let\r@fend=\endpaper\gdef\endpaper{\ifr@ffile
\immediate\write16{Cross References written on []\jobname.REF.}\fi\r@fend}

\catcode`@=12

\citeall\refto		
\citeall\ref		%
\citeall\Ref		%
\citeall\Refs		%

%
%
\elevenpoint\doublespace
\voffset 0.3in
\fontdimen13\twelvesy=5pt
\fontdimen14\twelvesy=5pt
\fontdimen15\twelvesy=5pt
\fontdimen16\twelvesy=5pt
\fontdimen17\twelvesy=5pt

\def\pb{\phantom{\bullet}\kern-3.5pt}

\title{Bosonization and Fermion Liquids in Dimensions Greater Than One}
\bigskip
\author{A. Houghton}
\affil{
Department of Physics
Box 1843
Brown University
Providence, RI 02912 U.S.A.}
\author{and}
\author{J. B. Marston}
\affil{
Institute for Theoretical Physics
University of California
Santa Barbara, CA 93106-4030
and
Department of Physics
Box 1843
Brown University
Providence, RI 02912 U.S.A.}
\bigskip\bigskip\bigskip
\centerline{October 21, 1992}

\vfill\eject

\centerline{}

\vskip 1in

\abstract{We develop and describe new approaches to the problem of
interacting Fermions in spatial dimensions greater than one.  These
approaches are based on generalizations of powerful tools previously
applied to problems in one spatial dimension.  We begin with
a review of one-dimensional interacting Fermions.  We then introduce
a simplified model in two spatial dimensions to study the role that spin
and perfect nesting play in destabilizing Fermion liquids. The complicated
functional renormalization group equations of the full problem
are made tractable in our
model by replacing the continuum of points that make up the closed
Fermi line with four Fermi points.  Despite this
drastic approximation, the model exhibits physically reasonable
behavior both at half-filling (where instabilities occur)
and away from half-filling (where a Luttinger liquid arises).
Next we implement the Bosonization of higher dimensional
Fermi surfaces introduced by Luther and advocated most recently by Haldane.
Bosonization incorporates the phase space and small-angle scattering
processes neglected in our model (but does not, as yet, address
questions of stability).  The charge sector is equivalent to an
exactly solvable Gaussian quantum
field theory; the spin sector, however, must be solved semiclassically.
Using the Luther-Haldane approach we recover the collective mode equation
of Fermi-liquid theory and in three dimensions reproduce
the $T^3~ {\rm ln}(T)$ contribution to the specific heat due to
small angle scattering processes.  We conclude with a discussion of our
results and some speculation about future possibilities.}

\pacs{03.65.Fd, 11.40.Dw, 71.10.+x, 71.45.-d}

\endtitlepage

\centerline{I. INTRODUCTION AND REVIEW OF A 1-D MODEL WITH 2-FOLD U(1)
SYMMETRY}
\smallskip
Fermi liquid theory is now nearly forty years old.  It is important
to ascertain the range of its validity and determine whether more exotic
generalizations, such as Luttinger liquids which exhibit
spin-charge separation, exist.
Shankar recently has emphasized the advantage
of the renormalization-group (RG) approach over various types of mean-field
approximations for answering these questions\refto{Shankar1}.  In short,
mean-field descriptions prejudice the outcome of the analysis
by assuming that one,
or at most a few, type of instability dominate the physics.  RG
analysis, on the other hand, treats all possible instabilities on an
equal footing.  Unfortunately, in spatial dimensions greater than one,
the RG flows are described by nearly intractable functional equations.

The approach we take in this paper is to slowly work up to the full
problem by first reviewing rather well known one dimensional
physics\refto{Emery}.
We then study a simplified model in two dimensions that incorporates some of
the
new physics that arises in higher spatial dimensions while still retaining
the simplicity of one dimensional systems.  Of course the price we pay
for this simplicity is the drastic approximation to physical reality
that we must make in order to arrive at the model:  we completely neglect
small angle scattering processes.
Nevertheless, the model suggests a way to
completely reformulate Fermi liquid theory.  Following Haldane's
suggestion\refto{Haldane0} we
now view Fermi and Luttinger liquids as zero temperature quantum critical
fixed points characterized by infinite U(1) symmetry.
The reformulation sheds light on how one
might go beyond the drastic approximations of the model
to include small angle scattering processes.

Spin-charge separation occurs automatically in one spatial dimension,
at least in the weak coupling limit and at long length scales.
Consider the following low energy
effective theory for excitations near the two Fermi points depicted in
Figure [1].  The action in the non-interacting limit is given by:
$$S_0 = \int dx~ dt~ \{ \psi_L^{\dagger \alpha} \partial_{-} \psi_{L \alpha}
+ \psi_R^{\dagger \alpha} \partial_{+} \psi_{R \alpha} \}\ .\eqno(1.1)$$
Here L and R refer to the left and right Fermi points; $\partial_{\pm}
\equiv \partial_{t} \mp i v_f \partial_{x} $
where $v_f$ is the Fermi velocity which we
will set equal to one for now. The Lorentz symmetry of this action
guarantees that the left moving Fermi fields are purely functions
of the combination $(x + i v_f t)$ whereas the right fields are
functions of $(x - i v_f t)$.
The electron destruction
fields $c_{\alpha}(x, t)$, where $\alpha = \uparrow$ or $\downarrow$
for up and down
spins (with summation convention assumed), are related to the slowly varying
continuum fields  $\psi_{L,R}$ by:
$$c_{\alpha}(x, t) \equiv {{1}\over{\sqrt{2}}}~ \{
e^{-i k_f x}~ \psi_{L \alpha}(x, t)
+ e^{i k_f x}~ \psi_{R \alpha}(x, t) \}\ .\eqno(1.2)$$
Upon substituting this form into any given microscopic Hamiltonian
the many-body interactions take the form of quartic and
higher powers of the continuum fields.
Most of the interactions are irrelevant in
the renormalization group sense (see below) and the most general
{\it marginal} interaction takes the form:
$$\eqalign{S_{int} &= \int dx~ dt~ \{ {{\pi}\over{2}}~
\delta v_c~ (J_L^2 + J_R^2)
+ {{\pi}\over{6}}~ \delta v_s~ (J_{L \alpha}^{\beta} J_{L \beta}^{\alpha}
+ J_{R \alpha}^{\beta} J_{R \beta}^{\alpha}) \cr
&+ \lambda_c~ J_L J_R
+ \lambda_s~ J_{L \alpha}^{\beta} J_{R \beta}^{\alpha} \}
. \cr} \eqno(1.3)$$
Here for instance the charge current at the left point is defined by
$J_L \equiv :\psi_L^{\dagger \alpha} \psi_{L \alpha}:$ where the normal
ordering symbols ``:'' indicate that we have subtracted the constant
background charge density from the current to make $<J_L (x, t)> = 0$.
The spin current is most conveniently expressed in matrix form:
$J^{\alpha}_{L \beta}(x) \equiv \psi_L^{\dagger \alpha}(x)~ \psi_{L \beta}(x)
- \half \delta^{\alpha}_{\beta}~ \psi_L^{\dagger \gamma}(x) \psi_{L
\gamma}(x)$.  The matrix form can always be converted into the more
familiar vector form with the identity:  $J^a_L = \half
(\sigma^a)^\beta_\alpha~ J^\alpha_{L \beta}$ where a = x, y, or z.
Note that the spin current
has no charge current component because it is traceless.  It also has
zero vacuum expectation value because in 1+1 dimensions
the vacuum cannot break the continuous SU(2)
spin rotational invariance by a quantum
generalization of the Mermin-Wagner theorem.
In the Fermion action, spin-charge separation is apparent
even before Bosonization. That is, the interaction term
involves only products of either pure spin or pure charge currents.
The Gaussian part of the action, $S_0$, also
can be expressed purely in terms of separate products of
the charge and spin currents (see below).

Omitted from the action are terms that oscillate
rapidly with wavevectors of order $k_f$, interactions involving
derivatives that arise from Taylor expansions of non-local interactions, and
terms with more than four Fermion fields.
Many of these terms break spin-charge separation; however, each is
irrelevant in the renormalization group sense and
the coupling constants flow rapidly to zero in the
low-energy limit.  To show the irrelevance, consider the scale transformation
$x \rightarrow s x$ and $t \rightarrow s t$ where $s > 1$.  The Gaussian
part of the action, $S_0$, remains invariant if we rescale the fields
$(\psi^\dagger, \psi) \rightarrow s^{-1/2} (\psi^\dagger, \psi)$.
(Note that any non-linearities in the dispersion relation due to band
structure are smoothed out as $s \rightarrow \infty$.)
Similarly, $S_{int}$ remains invariant, showing that it is a marginal
interaction.   All other terms will, however, scale away at least as
fast as an inverse power of $s$ when
$s \rightarrow \infty$.
Thus, in one dimension, spin-charge separation
occurs in the low-energy
effective theory regardless of how the marginal interactions flow.
Note that non-zero temperature acts as an infrared
cutoff (since the time direction has a finite extent $\beta \equiv
{{1}\over{k_B T}}$)
that stops scaling towards the low-energy region
beyond this scale.  Irrelevant terms therefore persist at non-zero
temperature so the phenomenon of spin-charge
separation must be construed as a zero-temperature
critical property of the theory.

Apart from the observation of spin-charge separation,
the low-energy theory can be classified in terms of the symmetries
that it obeys.  In addition to the global SU(2) spin
rotational symmetry, there exist two separate U(1) symmetries: one
for each Fermi point.  This U(1)$_L$ $\otimes$ U(1)$_R$ symmetry may
be exhibited by considering the effect of separate left and right phase
rotations by angles $\Gamma_L$ and $\Gamma_R$ on the Fermion variables:
$$\eqalign{
\psi_{L \alpha}(x, t) &\rightarrow e^{i \Gamma_L}~ \psi_{L, \alpha}
(x, t) \cr
\psi_{R \alpha}(x, t) &\rightarrow e^{i \Gamma_R}~ \psi_{R, \alpha}
(x, t) \ .\cr} \eqno(1.4)$$
All of the currents are clearly invariant under this transformation, as
the $\psi^\dagger$ fields transform with opposite phases.  The
physical meaning of the invariance is clear: the action, as it stands,
conserves separately the number of left and right particles.  We
shall see that this special property has a natural generalization
to higher spatial dimensions.
Actually, one other marginal four-Fermi interaction
can appear.  The Umklapp term
$$\lambda_3~ [(\psi_R^{\dagger \alpha} \psi_{L \alpha})^2 + H.c.] \eqno(1.5)$$
is permitted at half-filling in a periodic one-dimensional solid and it
breaks the $U(1)_L \otimes U(1)_R$ symmetry down to
the diagonal subgroup of ordinary $U(1)$ transformations
with $\Gamma_L = \Gamma_R$.
It violates the separate left and right U(1) symmetries because it
transports two particles from one Fermi point to the other.  Of course,
total particle number remains conserved, and this conservation is
reflected in the remaining diagonal U(1) symmetry.  Like the
other terms in the action, the Umklapp term
preserves spin-charge separation because
it transports charge, not spin, from one Fermi point to the other.  To see
this, note that it may be rewritten as:
$\half \lambda_3~ (\epsilon_{\alpha
\beta} \psi_R^{\dagger \alpha} \psi_R^{\dagger \beta}) (\epsilon^{\gamma
\delta} \psi_{L \gamma} \psi_{L \delta}) + H.c.$  where $\epsilon_{\alpha
\beta}$ is the totally antisymmetric tensor with $\epsilon_{12} = 1$.
Thus, only spin-singlet objects move from one Fermi point to the other.

We now return to the problem without the Umklapp term and determine the RG
flows and the nature of the fixed points.  Bosonization of the Fermion
fields is a powerful tool for addressing these questions.
For now we use Abelian Bosonization\refto{Shankar2}
and choose the spin quantization
axis in the $\bf \hat{z}$ direction.  The current algebra will provide
the vital link between the Fermion and Boson representations.  We start by
defining the normal-ordering operation carefully:
$$\eqalign{J_{L \alpha}(x, t)
&\equiv~ : \psi^\dagger_\alpha(x, t)~ \psi_\alpha(x + \epsilon, t) : \cr
&\equiv \psi^\dagger_\alpha(x, t)~ \psi_\alpha(x + \epsilon, t)
- < \psi^\dagger_\alpha(x, t)~ \psi_\alpha(x + \epsilon, t) >\
.\cr}\eqno(1.6)$$
Here we place the spin index as a subscript on the $\psi^\dagger$ field
to emphasize that we
are no longer summing over it, and we imagine taking the $\epsilon \rightarrow
0$ limit at the end of our calculations.  This ``point-splitting'' procedure
regularizes ultraviolet divergences in our calculation.  We choose real-space
regularization because the connection between Bosons and Fermions occurs most
naturally in real space.  Momentum space regularization will be introduced
later to permit the evaluation of momentum space integrals; differences
between the
two regularization procedures do not change the low-energy results.
Currents for the right moving
sector are obtained by making the replacement $L \rightarrow R$
and a simple calculation shows that left currents commute with right currents,
while two left or two right currents at equal times obey the Kac-Moody algebra:
$$[J_{L \alpha}(x)~ ,~ J_{L \beta}(y)] = -{{i}\over{2 \pi}}~
\delta_{\alpha \beta}~ \delta'(x - y) \ .$$
$$[J_{R \alpha}(x)~ ,~ J_{R \beta}(y)] = +{{i}\over{2 \pi}}~
\delta_{\alpha \beta}~ \delta'(x - y) \ .\eqno(1.7)$$
(To derive these relations, use the equal-time propagators for the Fermions
$<\psi^{\dagger \alpha}_L(x) \psi_{L \beta}(0)> = \delta^\alpha_\beta~
{{-i}\over{2 \pi x}}$ and
$<\psi^{\dagger \alpha}_R(x) \psi_{R \beta}(0)> = \delta^\alpha_\beta~
{{+i}\over{2 \pi x}}$.)
The coefficient of the derivative of the Dirac $\delta$-function is
known as the quantum anomaly.  Note that it has the opposite sign
for the left versus the right movers.  The charge current defined previously
may now be expressed in terms of these currents
as: $J_L(x) = J_{L \uparrow}(x) +
J_{L \downarrow}(x)$ and the $z$ component of the spin current is
simply: $J_{L z}(x) = J_{L \uparrow}(x) - J_{L \downarrow}(x)$.
(However, the other two components of the spin current $J_{L x}$
and $J_{L y}$ are not so simply related to $J_{L \uparrow}$ and
$J_{L \downarrow}$ .)
The charge and spin currents also obey the Kac-Moody algebra,
but with twice the anomaly.

We now introduce real-valued left and right moving free Boson fields
$\phi_{L \alpha}$ and $\phi_{R \alpha}$
which satisfy the commutation relations:
$$[\phi_{L \alpha}(x)~ ,~ \phi_{L \beta}(y)] = -{{i}\over{4}}~
\epsilon(x - y)\ ,$$
$$[\phi_{R \alpha}(x)~ ,~ \phi_{R \beta}(y)] = +{{i}\over{4}}~
\epsilon(x - y)\ ,\eqno(1.8)$$
and
$$[\phi_{R \alpha}(x)~ ,~ \phi_{L \beta}(y)] = i/4\ , \eqno(1.9)$$
where $\epsilon(x) = 1$ for $x > 0$ and $= -1$ for $x < 0$.
We also define canonical Boson currents
$$J_{L \alpha}(x, t) \equiv
-{{1}\over{\sqrt{\pi}}}~ {{\partial \phi_{L \alpha}(x, t)}\over{\partial x}} $$
and
$$J_{R \alpha}(x, t) \equiv +{{1}\over{\sqrt{\pi}}}~
{{\partial \phi_{R \alpha}(x, t)}\over{\partial x}}\ .\eqno(1.10)$$
It is a remarkable fact that these Boson currents obey the same Kac-Moody
algebra as the Fermion currents defined previously in Eq. [1.6].
To check the current commutation relation, take spatial derivatives of the
Free Boson propagators:
$$< \phi_{L \alpha}(x) \phi_{L \alpha}(0) - \phi^2_{L \alpha}(0) >
= {{1}\over{4 \pi}}~ {\rm ln}~ {{a}\over{a + ix}}$$
$$< \phi_{R \alpha}(x) \phi_{R \alpha}(0) - \phi^2_{R \alpha}(0) >
= {{1}\over{4 \pi}}~ {\rm ln}~ {{a}\over{a - ix}} \eqno(1.11)$$
(no sum over $\alpha$ in these expressions)
to form the expectation value of the commutator.
Here we see the first indication of equivalence between the Fermion and
Boson representations.
A further connection is revealed by examining the spectra of the free
Fermion and Boson theories: both of which are linear and are
in fact identical.
To go further, we rewrite the quadratic Boson Hamiltonian
in terms of the Boson currents defined in Eq. [1.10]:
$$H = \pi~\sum_{\alpha = \uparrow, \downarrow}~ \int dx~
\{J_{L \alpha}^2(x) + J_{R \alpha}^2(x) \}\ .\eqno(1.12)$$
The key point to be made here is that the currents appearing in this
Hamiltonian could just as
well be the Fermion currents Eq. [1.6] since the algebras are identical.
Although in this representation the Hamiltonian is a quartic function
of the Fermion fields and as such would appear to be intractable,
remarkably, as we have seen, it is equivalent to a free
Fermion theory.

We now consider the effect of the three types of interactions appearing
in the action Eq. [1.3].  First,
the current bilinears proportional to $\delta v_c$ and $\delta v_s$
simply renormalize the coefficients of the quadratic
Boson Hamiltonian.  [Although all three spin components of the spin current
appear in Eq. [1.3], SU(2) invariance means that $J^\alpha_{L \beta}(x)
J^\beta_{L \alpha}(x)$ can be replaced by $3 J_{L z}^2 (x)$ so in this case
it suffices to consider only
the z-component of the spin current.]  Using:
$$J_{L \uparrow}^2 + J_{L \downarrow}^2 = \half \{ J_L^2 + J_{L z}^2 \}
\eqno(1.13)$$
we see that so far spin-charge separation is explicit as the renormalized
Boson Hamiltonian may now
be expressed as the sum of two pieces, $H = H_c + H_s$, that separately
describe charge and spin excitations propagating at different velocities:
$$H_c = {{\pi}\over{2}}~ v_c~ \int dx~ \{J_L^2(x) + J_R^2(x) \} \eqno(1.14)$$
and
$$H_s = {{\pi}\over{2}}~ v_s~ \int dx~ \{J_{L z}^2(x) + J_{R z}^2(x) \}
\eqno(1.15)$$
where $v_c = 1 + \delta v_c$ and $v_s = 1 + \delta v_s$.
It should be emphasized
that the coefficients $\delta v_c$ and $\delta v_s$ do not
flow under RG transformations but instead simply renormalize the bare
velocities.  It may seem strange to have
two different ``velocities of light'' -- indeed, Lorentz invariance is
broken.  But since the charge and spin sectors are separate, Lorentz
invariance is now manifest separately in the two sectors.

Similarly, the charge current coupling, $\lambda_c$, in the action
Eq. [1.3] remains fixed
as $s \rightarrow \infty$.  To see this, note that it may be incorporated
by adding another term quadratic in the Boson currents
to the charge Hamiltonian:
$$H_c = \int dx~ \{{{\pi}\over{2}}~ [J_L^2(x) + J_R^2(x)]
+ \lambda_c~ J_L(x)~ J_R(x) \}\ . \eqno(1.16)$$
To determine the effect of $\lambda_c$ on the spectrum, we must diagonalize
this Hamiltonian.  Diagonalization is accomplished via a
Bogoliubov transformation that respects the Kac-Moody algebra.
Let:
$$J_L(x) = {\rm cosh}(\eta)~ J'_L(x) + {\rm sinh}(\eta)~ J'_R(x)$$
and
$$J_R(x) = {\rm sinh}(\eta)~ J'_L(x) + {\rm cosh}(\eta)~ J'_R(x)\
.\eqno(1.17)$$
Then the primed charge currents obey the same algebra as the original
(unprimed) charge currents.  Upon substituting these currents into the
Hamiltonian Eq. [1.16] we find that the choice
$${\rm tanh}(2 \eta) = - {{\lambda_c}\over{\pi}} \eqno(1.18)$$
eliminates cross terms of the form $J'_L(x)~ J'_R(x)$.
We now introduce primed Boson fields $\phi'_{L c}(x) \equiv
{{1}\over{\sqrt{2}}}~ (\phi'_{L \uparrow} + \phi'_{L \downarrow})$ and
$\phi'_{R c}$ associated with the primed charge currents.  A factor of
$\sqrt{2}$ is needed to reproduce the correct anomaly:
$J'_L(x) \equiv \sqrt{2/\pi}~ \partial_x \phi'_{L c}(x)$
with a similar formula for the right sector.  The Hamiltonian
written in terms of these fields is simply:
$$H_c = (1 - {{\lambda_c^2}\over{\pi^2}})^\half~
\int dx~ {\Bigl \lbrace} ({{\partial \phi'_{L c}}\over{\partial x}})^2
+ ({{\partial \phi'_{R c}}\over{\partial x}})^2 {\Bigr \rbrace}\ .\eqno(1.19)$$
Thus the Bosonic theory remains Gaussian, even for $\lambda_c \neq 0$.

It might be expected that the spin current coupling $\lambda_s$
could be incorporated in a similar fashion.
However, the spin interaction
$\lambda_s~ J^\alpha_{L \beta}(x)~ J^\beta_{R \alpha}(x)$ differs in a
fundamental way from the charge interaction $\lambda_c~ J_L(x)~ J_R(x)$.
Only the $J_{L z} J_{L z}$ part of the interaction
has a quadratic representation
in terms of the Boson field $\phi_s$.  The other two components are
rather more complicated.  Note that SU(2) invariance may be employed
only when both currents in the bilinear are of the left or right type;
for example, when we replace
$J^\alpha_{L \beta}(x) J^\beta_{L \alpha}(x) \rightarrow 3 J_{L z}^2 (x)$.
The non-trivial nature of this term is apparent in the Fermion basis:
$\lambda_s$ is the only interaction coefficient in $S_{int}$ that renormalizes.
To second order in weak-coupling perturbation theory (see below) it flows as:
$${{d\lambda_s}\over{d({\rm ln}(s))}} = 2 \pi (\lambda_s)^2\ . \eqno(1.20)$$
Because no fixed points intervene
at intermediate coupling in the original lattice Hubbard model (which
was exactly solved by Lieb and Wu via the Bethe ansatz\refto{LW}) the flow
for the continuum problem described by Eq. [1.20] is likely
to be qualitatively correct at {\it all} $\lambda_s$.
The $\lambda_s = 0$ fixed point of Eq. [1.20] (which is stable
when approached from the $\lambda_s < 0$ side) exhibits an enlarged
$SU(2)_L \otimes SU(2)_R$ symmetry because the left and right spin currents
are decoupled.  In other words, separate SU(2) rotations on the left and
right currents leave the fixed point action invariant.
The strong coupling fixed point, with $\lambda_s \rightarrow
\infty$ shows no such symmetry; instead a spin gap opens and the
electrons pair into singlets that require energy to break apart.

To understand better the role of spin-charge separation in the
one-dimensional problem, we examine the single-particle Green's
function along the fixed line $\lambda_s = 0$.  Following
Shankar\refto{Shankar2},
we introduce momentum space regularization by including a convergence
factor $e^{- \half a |p|}~ dp$ along with the integration measure, and
we take the $a \rightarrow 0$ limit at the end.
Now the (equal time)
correlation functions for the $\phi'_{L c} \equiv {{1}\over{\sqrt{2}}}~
[\phi'_{L \uparrow} + \phi'_{L \downarrow}]$ fields are given by:
$$\eqalign{G_{L c}(x) &\equiv
< \phi'_{L c}(x) \phi'_{L c}(0) - \phi'^2_{L c}(0) > \cr
&= {{1}\over{4 \pi}}~ {\rm ln}~ {{a}\over{a + ix}} \cr}$$
$$\eqalign{G_{R c}(x) &\equiv
< \phi'_{R c}(x) \phi'_{R c}(0) - \phi'^2_{R c}(0) > \cr
&= {{1}\over{4 \pi}}~ {\rm ln}~ {{a}\over{a - ix}}\ .\cr} \eqno(1.21)$$
The spin Bosons $\phi_s$ exhibit identical correlations.
It is easy to restore the time-dependence of these correlation
functions by using Lorentz invariance (of course with different velocities
in the charge and spin sectors).
The Bosonization procedure is completed with the observation that
Fermion operators are equivalent to exponentials of the original
unprimed Boson fields:
$$\eqalign{
\psi_{L \alpha} &= {{1}\over{\sqrt{2 \pi a}}}~
{\rm exp}[-i \sqrt{4\pi}~ \phi_{L \alpha}] \cr
\psi_{R \alpha} &= {{1}\over{\sqrt{2 \pi a}}}~
{\rm exp}[i \sqrt{4\pi}~ \phi_{R \alpha}]\ .\cr}
\eqno(1.22)$$
One way to prove relations Eq. [1.22] is by constructing the Fermion currents
with the point splitting procedure described above.
Then spatial derivatives of the
Boson fields appear, and the currents defined in Eq. [1.10] are obtained.
Since $\phi_{L c} = {\rm cosh}(\eta) \phi'_{L c} + {\rm sinh}(\eta) \phi'_{R
c}$
and $\phi_{L \uparrow} = {{1}\over{\sqrt{2}}}~
(\phi_{L c} + \phi_{L s})$ etc. we
can combine the Bosonization formulas to find the Fermion two-point functions.
$$\eqalign{<\psi_L^{\dagger \beta}(x, t)~ \psi_{L \beta}(0, 0)>
&= 2 \times
{{1}\over{2 \pi a}}~ {\rm exp}\{2 \pi [{\rm cosh}^2(\eta)~ G_{L c}(x, t)
+ {\rm sinh}^2(\eta)~ G_{R c}(x, t)] \} \cr
&\ \ \ \ \times {\rm exp}\{2 \pi G_{L s}(x, t) \} \cr
&= {{1}\over{\pi}}~ (x - i v_s t)^{-1/2}
(x - i v_c t)^{-1/2} (x^2 + v_c^2 t^2)^{- \alpha} \cr}$$
and
$$<\psi_R^{\dagger \beta}(x, t)~ \psi_{R \beta}(0, 0)> = {{1}\over{\pi}}~
(x + i v_s t)^{-1/2}
(x + i v_c t)^{-1/2} (x^2 + v_c^2 t^2)^{- \alpha}\ .\eqno(1.23)$$
Here the anomalous exponent $\alpha \equiv {\rm sinh}^2(\eta)$.
The explicit separation of charge
and spin reflects both the different velocities of the two types of
excitations and the remaining interaction $\lambda_c$ in the charge sector.

The path integral picture yields the following free Lagrangian
densities\refto{AM}:
$$\eqalign{
{\sl L}_c[\phi_c] &= \half~ (1 - {{2 \lambda_c}\over{\pi}}) \partial_{\mu}
\phi_c \partial^{\mu} \phi_c \cr
{\sl L}_s[\phi_s] &= \half~ \partial_{\mu}
\phi_s \partial^{\mu} \phi_s \cr}
\eqno(1.24)$$ where the spatial derivatives in these two expressions
implicitly include the different velocities factors, $v_c$ and $v_s$.
The Bosonization formulas Eq. [1.22] imply that the $U(1)_L
\otimes U(1)_R$ symmetry operation is effected simply by shifting the
left and right Bosons by, in general, different constants.  Thus,
$\phi_c$ is shifted
($\ \phi_{L c} \rightarrow \phi_{L c} + {{\Gamma_L}\over{\sqrt{2 \pi}}}$ and
$\phi_{R c} \rightarrow \phi_{R c} + {{\Gamma_R}\over{\sqrt{2 \pi}}}\ $ ),
but $\phi_s$ remains invariant, reflecting the
fact that the symmetry
operation acts only on the charge sector.  The Lagrangian density Eq. [1.24]
remains invariant because $\partial_\mu \phi_c$ is unaffected by the
shift.
The Umklapp term, as expected, breaks the symmetry because it is
equivalent to adding the term $\lambda_3~ {\rm cos}(\sqrt{8 \pi}~
\phi_c)$ to $L_c[\phi_c]$ and the cosine clearly changes under a shift
of $\phi_c$ by a constant.

At this point we might question whether the
strange form of the propagator somehow eliminates the logarithmic
divergences that give rise to the RG flow described by Eq. [1.20].  The
answer, which is no, may be seen easily in the Boson basis where
spin-charge decoupling is explicit.  There the interaction appears as a
term proportional to
$\lambda_s~ {\rm cos}(\sqrt{8\pi}~ \phi_s)$ added to $L_s[\phi_s]$
of Eq. [1.24] which drives the RG flow.
[Note that the $SU(2)_L \otimes
SU(2)_R$ symmetry at $\lambda_s = 0$ now manifests itself in the Boson basis
as separate shifts by a constant in $\phi_{L s}$ and $\phi_{R s}$.  Actually,
full SU(2) symmetry would exhibit invariance under three types of rotations
corresponding to the three generators of SU(2).  Abelian Bosonization,
however, forces us to choose a spin quantization axis; consequently the theory
only exhibits explicit symmetry under rotations about that axis.  Invariance
under rotations in the other two directions remains hidden.]
In the Fermion basis, the problem
is slightly more complicated.  The logarithmic divergence that drives the
flow described by Eq. [1.20] comes
from a single loop diagram with four external fields
that contains two propagators: one for left moving fields
and one for right movers (see Figure [2]).  (Diagrams with two left or
two right propagators do not yield logarithmic divergences.)
The diagram is most easily evaluated in position space.  The
integral to be evaluated is:
$$I(s) = (\lambda_s)^2~ \int^s_1 dx~ \int^\infty_{-\infty}
dt~ (x^2 + v_c^2 t^2)^{-\half - 2 \alpha}
(x^2 + v_s^2 t^2)^{-1/2}\ .\eqno(1.25)$$
Since we are performing a perturbative expansion to order $(\lambda_s)^2$
it is sufficient to set $\alpha = ({{\lambda_c}\over{2 \pi}})^2 +
O(\lambda_c^4)$
equal to zero and study whether the logarithmic singularity
persists when $v_c \neq v_s$.  With $\alpha = 0$ the spatial integral yields a
hypergeometric function which, when integrated over time, indeed produces the
desired logarithm.

Fermi liquid behavior arises only in the special case $v_c = v_s$ and
$\alpha = 0$ as the two-point function Eq. [1.23]
must have a single simple pole.  However, the
discontinuity at the Fermi surface remains even when $v_c \neq v_s$
so long as $\alpha = 0$.  To
see this, note that the momentum space occupancy is found by taking the
Fourier transform of the equal time propagator (which we
define to be the average of the correlation function evaluated
at times $t = 0+$ and $t = 0-$ in order to specify definite
time orderings which also cancel out the imaginary component).
Use of the Fermion anticommutation relations then shows:
$$2 n_L(k) - 1
= \half~ \int dx~ e^{i (k + k_f) x}~ \{
< \psi^{\dagger \beta}_L(x,t=0+) \psi_{L \beta}(0,0) >
+ < \psi^{\dagger \beta}_L(x,t=0-) \psi_{L \beta}(0,0) > \}\ .\eqno(1.26)$$
But as $t \rightarrow 0\pm$ the two velocities disappear from the
correlation function, which equals ${{|x|^{-2 \alpha}}\over{\pi x}}$ and yields
a step function in momentum space only for $\alpha = 0$.  Apparently
spin-charge separation and the destruction of the Fermi discontinuity
are separate issues.\refto{Ian}
Both are characteristic properties of Luttinger liquids\refto{Haldane2}.
In the following we continue to speak of Fermi points and Fermi surfaces.
Clearly these points or surfaces should now be defined more
generally as manifolds of points in momentum space at which
the zero-temperature
occupancy shows non-analytic behavior characterized by an exponent
$\alpha$ instead of a discontinuity.  In particular, near the Fermi
momentum the occupancy varies as: $n(k) \approx n(k_f) + C |k - k_f|^{2 \alpha}
sgn(k - k_f)$ where C is a non-universal constant that sets the
momentum scale.  It depends on the
the momentum-energy cutoff in the interaction $\lambda_c$.

The outline for the rest of the paper is as follows.  In section (II)
we introduce a model with four Fermi points in two spatial dimensions
that incorporates some key features of the higher dimension interacting
Fermion problem.  A renormalization group solution of the model
away from half-filling
finds a stable fixed line with four-fold U(1) symmetry that naturally
generalizes the one-dimensional $U(1)_L \otimes U(1)_R$ fixed point symmetry.
Bosonization of the Fermions at these points suggests a new way to look
at Fermions in higher dimension and in section (III) we follow this line
of thought to arrive at full Fermi surface fixed point manifolds with infinite
U(1) symmetry.  In section (IV) we demonstrate that this way of looking
at things yields concrete results by calculating non-analytic
contributions to the specific heat.  And in section (V) we rederive the
collective mode equation in the new framework.
Finally, section (VI) contains some
discussion and speculations.

\vfill\eject
\centerline{II. A 2-D FIXED POINT WITH 4-FOLD U(1) SYMMETRY}
\smallskip
We turn now to a simple model for two-dimensional interacting Fermions
that illustrates how possibly spin-charge separation might occur in
spatial dimensions greater than one.  The Fermi surface of a nearest
neighbor tight-binding model on a square lattice serves as inspiration
for the model.  Instead of treating the entire continuum of Fermi points that
make up the Fermi line enclosing the occupied states, we make our
problem tractable with the following drastic approximation: we treat
each of the four sides of the Fermi surface as a single point
labeled by $\pm 1$ or $\pm 2$ (see
Figure [3]).  At half-filling, these points lie respectively at momenta
$\pm(\pi/2, \pi/2)$ and $\pm (\pi/2, -\pi/2)$ but away from half-filling
the momentum is generally incommensurate with the reciprocal lattice
vectors.  With this simplification, the infinite set of
renormalization group equations is reduced to a manageable finite set.

Note that this model differs from models studied earlier by
Schulz\refto{Schulz}
and Dzyaloshinskii\refto{Dzy} that
focused on the Van Hove singularities at the four corners $(0, \pm \pi)$
and $(\pm \pi, 0)$ of the half-filled
tight binding spectrum.  Our model also is not equivalent to
two coupled parallel
Hubbard chains -- a system studied by Anderson\refto{PWA1},
Finkel'stein\refto{Fink}, and others.
It differs in that excitations at points $\pm 2$
propagate at right angles in momentum space with respect to
excitations at points $\pm 1$ whereas excitations in the two-chain system
always propagate in parallel (or anti-parallel) directions.  Consequently,
different marginal interactions and renormalization group equations appear
in our model.  Anderson's analysis\refto{PWA1}
of the two-chain problem led him to
conclude that weak interchain coupling does not change the one-dimensional
physics significantly but recent work by Finkel'stein\refto{Fink}
and Castellani, Di Castro and Metzner\refto{Cast}
suggests that interchain coupling is relevant and destabilizes the
Luttinger liquid.

We choose the model in part because it emphasizes
the role that perfect nesting plays in
destabilizing various fixed points (we elaborate on the nature of these
fixed points below).  Thus for example processes that transfer
an electron across the Fermi surface
from, say, point 1 to point -1 receive the same weight as processes that move
an electron from point 1 to 2 because the density of states is
non-zero only at the four points.  Note that the density of states at
each point must be held constant regardless of the system size.
One might be tempted to give each point the
same weight as the entire line it replaces, but this choice
proves uninteresting as quantum fluctuations would be suppressed
in the thermodynamic limit.
Photoemission experiments on the cuprate superconductors\refto{Photo}
provide another justification for our model\refto{Shankar3}.
These experiments show that hole pockets
form around the momentum points $\pm (\pi/2, \pi/2)$ and $\pm (\pi/2, -\pi/2)$
as the compounds are doped away from the insulating limit.  Low-energy
excitations near these points may
play an important role in the normal and superconducting phases.

Obviously van Hove singularities are ignored in our model.
They break scale invariance in the Gaussian part of the action
and therefore cannot be incorporated into the renormalization group scheme
because the dispersion relation is not linear at those
points.  While the
singularities may or may not be important at half-filling,
the Umklapp terms drive various instabilities which the
singularities are unlikely to prevent.
In any case, here it is the problem away from half-filling that
is of most concern to us.
The other major limitation of our model -- no small
angle scattering processes  -- is a severe approximation to physical
reality.  Small angle
scattering is clearly important for example in the formation of
momentum-space Cooper pairs.  We therefore expect, and indeed find,
unphysical behavior in certain limits (see below).  Nevertheless,
our calculation suggests that stable fixed points describing
whole Fermi surfaces, not just points, exist.

The electron operators $c_{{\bf x} \alpha}$
may be written in terms of the continuum
fields at the four points.  We now allow $\alpha$ to take on
values 1, ..., n for the SU(n) case.  We consider the
general problem because it enables us to check our calculations more
thoroughly for combinatorial errors.  It also permits us to study the
spinless case n = 1.
Let $u \equiv x + y$ and $v \equiv x - y$ where
$x$ and $y$ are integers labeling the coordinates of a point on a
lattice with unit lattice constant and place the system in periodic box so
that $1 \leq u,~ v \leq L$.  Then at half-filling
the lattice electron annihilation
operator can be rewritten in terms of the continuum fields at the four
points as:
$$c_{{\bf x} \alpha} = \half \{
e^{i \pi u / 2}~ \psi_{1 \alpha}(u) + e^{-i \pi u / 2}~
\psi_{-1 \alpha}(u)
+ e^{i \pi v / 2}~ \psi_{2 \alpha}(v) + e^{-i \pi v / 2}~
\psi_{-2 \alpha}(v) \}\ .\eqno(2.1)$$
Away from half-filling we simply replace $\pi/2 \rightarrow k_f$ in this
formula.  With the replacement Eq. [2.1] we see
that excitations are constrained to move in directions perpendicular
to the Fermi ``edges''.  In other words, the Fermions cannot move in
arbitrary directions, just forward and backwards along the lines depicted
in Figure [3].
With this definition we can now break up any four Fermi interaction into
two types of terms: marginal terms that vary smoothly in space and irrelevant
terms that oscillate rapidly or contain derivatives.
Keeping track of just the marginal terms
a little algebra shows, for example, that at half-filling
the Hubbard interaction
${{U}\over{n}}~ (c_{\bf x}^{\dagger \alpha} c_{{\bf x} \alpha} - n/2)^2$
generates the following four Fermi terms:
$$\eqalign{
& (\psi_{-2}^{\dagger \alpha} \psi_{-2 \alpha}
+ \psi_{-1}^{\dagger \alpha} \psi_{-1 \alpha}
+ \psi_{1}^{\dagger \alpha} \psi_{1 \alpha}
+ \psi_{2}^{\dagger \alpha} \psi_{2 \alpha})^2 \cr
&+ (\psi_{2}^{\dagger \alpha} \psi_{-2 \alpha}
+ \psi_{1}^{\dagger \alpha} \psi_{-1 \alpha}
+ \psi_{-1}^{\dagger \alpha} \psi_{1 \alpha}
+ \psi_{-2}^{\dagger \alpha} \psi_{2 \alpha})^2 \cr
&+ (\psi_{-2}^{\dagger \alpha} \psi_{-1 \alpha}
+ \psi_{-1}^{\dagger \alpha} \psi_{-2 \alpha}
+ \psi_{2}^{\dagger \alpha} \psi_{1 \alpha}
+ \psi_{1}^{\dagger \alpha} \psi_{2 \alpha})^2 \cr
&+ (\psi_{1}^{\dagger \alpha} \psi_{-2 \alpha}
+ \psi_{-1}^{\dagger \alpha} \psi_{2 \alpha}
+ \psi_{2}^{\dagger \alpha} \psi_{-1 \alpha}
+ \psi_{-2}^{\dagger \alpha} \psi_{1 \alpha})^2 . \cr} \eqno(2.2) $$
More generally, we can write down all possible four Fermi interactions
consistent with the symmetries of the SU(n) spin group and the symmetry
of the square Fermi surface.  Thus the perturbation is:
$$\eqalign{H_{int} &= {{1}\over{L}}~ \int du~ dv~ {\Bigl \lbrace}
\lambda_{1c}~ (J_1 J_{-1} + J_2 J_{-2})
+ \lambda_{1s}~ (J^\alpha_{1 \beta} J^\beta_{-1 \alpha}
+ J^\alpha_{2 \beta} J^\beta_{-2 \alpha}) \cr
&+ \lambda_{2c}~ (J_1 + J_{-1}) (J_2 + J_{-2})
+ \lambda_{2s}~ (J^\alpha_{1 \beta} + J^\alpha_{-1 \beta})
(J^\beta_{2 \alpha} + J^\beta_{-2 \alpha}) \cr
&+ \lambda_3~ [(\psi_1^{\dagger \alpha} \psi_{-1 \alpha})^2
+ (\psi_2^{\dagger \alpha} \psi_{-2 \alpha})^2 + H.c.] \cr
&+ \lambda_4~ (\psi_1^{\dagger \alpha} \psi_{-2 \alpha}
\psi_2^{\dagger \beta} \psi_{-1 \beta}
+ \psi_{-2}^{\dagger \alpha} \psi_{-1 \alpha}
\psi_1^{\dagger \beta} \psi_{2 \beta} + H.c.) \cr
&+ \lambda_5~ (\psi_1^{\dagger \alpha} \psi_{-2 \alpha}
\psi_{-1}^{\dagger \beta} \psi_{2 \beta}
+ \psi_{-2}^{\dagger \alpha} \psi_{-1 \alpha}
\psi_2^{\dagger \beta} \psi_{1 \beta} + H.c.) \cr
&+ \lambda_6~ (\psi_1^{\dagger \alpha} \psi_{-1 \alpha}
\psi_{-2}^{\dagger \beta} \psi_{2 \beta}
+ \psi_{1}^{\dagger \alpha} \psi_{-1 \alpha}
\psi_2^{\dagger \beta} \psi_{-2 \beta} + H.c.) \cr
&+ \lambda_7~ [(\psi_1^{\dagger \alpha} \psi_{2 \alpha})^2 +
(\psi_{-2}^{\dagger \alpha} \psi_{-1 \alpha})^2
+ (\psi_{2}^{\dagger \alpha} \psi_{-1 \alpha})^2 +
(\psi_{-2}^{\dagger \alpha} \psi_{1 \alpha})^2 + H.c.] {\Bigr \rbrace}
. \cr} \eqno(2.3) $$
Here we have again introduced the charge and spin currents, now for each of the
four points.  Note that $J_{\pm 1} = J_{\pm 1}(u)$,
$J_{\pm 2} = J_{\pm 2}(v)$, and likewise for the spin currents.
We rescale the interactions by a factor of ${{1}\over{L}}$ to
keep the density of states constant.
Not included in the above expression are terms like ${{\pi}\over{2}}~
\delta v_c~ (J_1)^2$ and
${{\pi}\over{6}}~ \delta v_s~ J_{1 \alpha}^{\beta} J_{1 \beta}^{\alpha}$
which simply
renormalize the charge and spin velocities.  For the above on-site
Hubbard interaction, the coupling constants take the following values:
$$\eqalign{\lambda_{1c} &= \lambda_{2c} = (U/n)(2 - 2/n) \cr
\lambda_{1s} &= \lambda_{2s} = -2U/n \cr
\lambda_3 &= U/n \cr
\lambda_4 &= 2U/n \cr
\lambda_5 &= 2U/n \cr
\lambda_6 &= 2U/n \cr
\lambda_7 &= U/n \cr} \eqno(2.4) $$
but other bare interactions (ie. nearest-neighbor Coulomb or spin
exchange) yield other values (see below).

Each of the nine coupling constants
in Eq. [2.3] corresponds to a particular process
drawn in Figure [4].  Unlike the one dimensional case, we see that
a number of these marginal processes break spin-charge separation.  The
current-current terms $\lambda_{1c}$, $\lambda_{2c}$, $\lambda_{1s}$, and
$\lambda_{2s}$ respect it, and so do the Umklapp terms
$\lambda_3$ and $\lambda_7$ (at least for the physical SU(2) problem),
but the other terms ($\lambda_5$ and two of the Umklapp terms $\lambda_4$
and $\lambda_6$) are ``mixed'' processes that scatter both spin and charge.
Away from half-filling, the term $\lambda_5$ survives
and it is this interaction that will draw our attention in the following
renormalization group analysis.
As in one dimension, the model possesses global SU(2) spin symmetry.
However, it exhibits the four separate U(1) symmetries only if
$\lambda_3 = \lambda_4 = \lambda_5 = \lambda_6 = \lambda_7 = 0$.  In
other words, only current-current type interactions preserve
$U(1)_1 \otimes U(1)_{2} \otimes U(1)_{-1} \otimes U(1)_{-2}$ symmetry.
This behavior clearly differs from the one dimensional model, which
automatically exhibits $U(1)_L \otimes U(1)_R$ symmetry away from
half-filling.  Again $\lambda_5$ is the sole offending term away from
half-filling.

It is a straightforward, though lengthy, exercise to work out the RG
flows to second order in the coupling strengths by evaluating one loop
diagrams with four external Fermi field lines.  The diagrams are
essentially no different from the one we evaluated in one spatial
dimension (Figure [2]).  This is because at the one loop
level only diagrams that contain both a 1
propagator (ie. $< \psi_1^{\dagger \alpha}(u, t)~ \psi_{1 \alpha}(0, 0)
>$) and a -1 propagator (or 2 and -2 propagators)
yield logarithmically divergent contributions.
It follows that $\lambda_{2c}$, $\lambda_{2s}$, and
$\lambda_7$ do not flow at this order because these interactions contain
Fermions at points 1 and 2 (or -1 and 2, etc.) so the requisite propagators do
not appear.  Inspection of the diagrams in Figure [4] reveals the physical
origin for this decoupling.  The three interactions $\lambda_{2c}$,
$\lambda_{2s}$, and $\lambda_7$ differ from the other terms in that
exchange of momentum between the two points is forbidden because the
two directions are perpendicular.  For example,
consider the momentum-space version of interactions $\lambda_{1c}$ and
$\lambda_{2c}$.  Let $p$ denote momentum in the u-direction and
$q$ be momentum in the v-direction.  Then the interactions take the form:
$$\lambda_{1c}~ {\Bigl \lbrace}~ \int {{dp}\over{2 \pi}}~
J_1(p)~ J_{-1}(-p) + \int {{dq}\over{2 \pi}}~
J_2(q)~ J_{-2}(-q) {\Bigr \rbrace}$$ and
$$\lambda_{2c}~ L~
[J_1(p=0) + J_{-1}(p=0)] [J_2(q=0) + J_{-2}(q=0)]\ .\eqno(2.5)$$
In contrast to $\lambda_{1c}$ (and $\lambda_{1s}$),
only the zero-momentum component of
the currents couple in the $\lambda_{2c}$ (and $\lambda_{2s}$) terms.

The remaining six couplings
flow as follows [the prime denotes a derivative with respect to
$\pi {\rm ln}(s)\ $]:
$$\eqalign{
\lambda_{1c}' &= 8(1 - 1/n) (\lambda_3)^2 + 2 (\lambda_4)^2 - (2/n) \lambda_4
\lambda_6 + 2(1/n - 1) (\lambda_5)^2 + 2 (\lambda_6)^2 \cr
\lambda_{1s}' &= n (\lambda_{1s})^2 + 4(n - 2) (\lambda_3)^2 - 2 \lambda_4
\lambda_6 + 2 (\lambda_5)^2 + 2n (\lambda_6)^2 \cr
\lambda_3' &= 4 \lambda_{1c} \lambda_3 + 2(n - 1 - 2/n) \lambda_{1s}~
\lambda_3 + (\lambda_4)^2 + 2 \lambda_4 \lambda_6 - n (\lambda_6)^2 \cr
\lambda_4' &= [2 \lambda_{1c} - (2/n) \lambda_{1s} + 4 \lambda_3]
\lambda_4 \cr
\lambda_5' &= [-2 \lambda_{1c} + 2(1 + 1/n) \lambda_{1s}] \lambda_5 \cr
\lambda_6' &= [2 \lambda_{1c} + 2(n - 1/n) \lambda_{1s} + 4(1 - n)
\lambda_3] \lambda_6 + [4 \lambda_3 - 2 \lambda_{1s}] \lambda_4 \ .\cr}
\eqno(2.6)$$
We can perform several checks on these equations.  First, the equations
close: we have not forgotten any marginal operators.  Second,
if we consider
only the pair of points 1 and -1, the equations must reduce to the known
ones in one spatial dimension.  By setting $\lambda_4 = \lambda_5
= \lambda_6 = 0$ one can easily check that the remaining equations
($\lambda_3$ is now the Umklapp term mentioned in the previous section)
do agree with the known results in one dimension.
As another test, note that two
terms vanish in the physical case $n = 2$.  In particular,
$\lambda_3$ no longer couples to $\lambda_{1s}$ because the Umklapp term
is a SU(n) singlet operator {\it only} for the special SU(2) case.
Finally, what happens when n = 1, the case of spinless Fermions?
Many terms vanish because they do not exist for spinless Fermions.
Thus, the spin singlet Umklapp terms $\lambda_3 = \lambda_7 = 0$ by the
Pauli exclusion principle and of course the
spin current terms $\lambda_{1s} = \lambda_{2s} = 0$ because
there is no spin.  Also, $\lambda_6
= -\lambda_4$ because $\lambda_4$ processes can no
longer be distinguished from $\lambda_6$ ones and $\lambda_5 = 0$
due to internal cancellations present in Eq. [2.3] when the spin label
is removed.  Equations Eq. [2.6] respect this limit.

The RG flows described by Eq. [2.6] generically flow to large
values.  The flows are physically sensible: at half-filling Umklapp processes
generate various instabilities and the system becomes
gapped in the charge sector when the interactions are repulsive.
Attractive interactions, on the other hand can lead to superconducting
instabilities.  The restricted phase space of our model obscures
the interpretation of these instabilities.  For example,
the Goldstone mechanism tells us that phases of broken SU(n) symmetry
exhibit gapless spin excitations.  On the other hand, the Higgs mechanism
suppresses gapless excitations in the charge sector if the U(1) symmetry
breaks.  But our model retains the character of 1+1
dimensional phase space which is not large enough
to foster broken continuous symmetries.
In any case, our failure to treat the van Hove
singularities and small angle scattering processes is not as important
as it might first seem: these processes are unlikely to inhibit the formation
of instabilities.

The spinless case n = 1 is an exception.  As noted above, we
can take $\lambda_{1s} = \lambda_3 = \lambda_5 = 0$ and $\lambda_4 =
- \lambda_6$ in this case.  The flows are described by the Kosterlitz
- Thouless equations:
$$\eqalign{
\lambda_{1c}' &= 6 (\lambda_6)^2 \cr
\lambda_6' &= 2 \lambda_{1c} \lambda_6 \ .\cr}
\eqno(2.7)$$
Here the fixed line $\lambda_6 = 0$ is stable for $\lambda_{1c} \leq 0$.
We may interpret the instability at positive $\lambda_{1c}$ as a
tendency to form a charge-density wave.  To see this, note that the
next-nearest-neighbor Hubbard repulsion ${{U_1}\over{n}} (n_{\bf x
+ \hat{x}} + n_{\bf x + \hat{y}})~ n_{\bf x}$, where $n_{\bf x} \equiv c_{\bf
x}^{\dagger \alpha} c_{{\bf x} \alpha}$ is the electron occupancy at
site {\bf x},
leads to the following bare continuum
couplings at half-filling:
$$\eqalign{\lambda_{1c} &= \lambda_{2c} = (U_1/n)(2 + 2/n) \cr
\lambda_{1s} &= \lambda_{2s} = 2U_1/n \cr
\lambda_3 &= -U_1/n \cr
\lambda_4 &= 0 \cr
\lambda_5 &= 0 \cr
\lambda_6 &= -2U_1/n \cr
\lambda_7 &= 0 \ .\cr} \eqno(2.8) $$
(For the case n = 1, $\lambda_3$ should be set equal to zero
by the Pauli exclusion principle.)
So repulsive nearest-neighbor interactions grow, a tendency towards
the formation of a charge-density-wave
sets in and sites on the even sublattice exhibit different charge
density than those on the odd sublattice.  This behavior is consistent
with that found by Shankar\refto{Shankar1}
in his functional RG calculation for spinless
Fermions and therefore lends credibility to our model.  On the other
hand, Shankar finds a superconducting instability for attractive
interactions which contrasts with the stability shown by our model.
We reconcile this difference by noting that the Cooper instability is
driven by small angle scattering
processes that scatter pairs of Fermions of opposite momentum
around the Fermi surface.  Again the phase space for such processes in our
model is severely restricted by existence of only four Fermi points, and
we should therefore not expect momentum-space Cooper pairs.  Real-space
Cooper pairs can arise, however, as we show below.  Nevertheless, negative
$\lambda_{1c}$ corresponds to an attractive interaction and is
indicative of a tendency towards the formation of superconducting pairs.

The problem of spinning Fermions away from half-filling is
rather more interesting.  Setting the Umklapp terms
$\lambda_3 = \lambda_4 = \lambda_6
= \lambda_7 = 0$ we obtain the reduced set of flow equations:
$$\eqalign{
\lambda_{1c}' &= 2(1/n - 1) (\lambda_5)^2 \cr
\lambda_{1s}' &= n (\lambda_{1s})^2 + 2 (\lambda_5)^2 \cr
\lambda_5' &= 2[(1 + 1/n) \lambda_{1s} - \lambda_{1c}]~ \lambda_5 \ .\cr}
\eqno(2.9)$$
Here we find a stable fixed line defined by $\lambda_{1s} = \lambda_{5} = 0$
and
$\lambda_{1c} \geq 0$.  In fact numerical integration shows that
it attracts flows starting from
the repulsive Hubbard couplings given by Eq. [2.4] (see Figure [5]).
Again, $\lambda_{2c}$ and $\lambda_{2s}$ do not
renormalize at second order.  Like $\lambda_{1c}$, these couplings can
be non-zero along the fixed line.
The instability at negative $\lambda_{1c}$
can be interpreted as a tendency to form real-space Cooper pairs
something like those proposed in the original resonating-valence-bond (RVB)
theory of Anderson\refto{PWA} and collaborators.  Thus the Fermion spin permits
the formation of singlet pairs.

What is the nature of the stable region of the fixed line?  First
note that it exhibits both spin-charge separation and four-fold
U(1) symmetry because only current-current type interactions
remain.  The fixed line thus represents a natural generalization of
the one-dimensional Luttinger liquid and as such motivates the approach
to the continuum Fermi surface problem we describe in the next
section.  Let us first look more closely at our
solution by transforming to Boson variables.  At first it seems strange
to contemplate Bosonization in spacetime dimensions greater than two.
But the problem remains essentially one-dimensional; the Fermions at
each of the four points are restricted to move along lines.  In fact
it is convenient to introduce complex space-time coordinates analogous
to those in 1 + 1 dimensions.  Let $u_{\pm} \equiv u \pm i t$ and
$v_{\pm} \equiv v \pm i t$ where we remember that the velocities, here set
equal to one, are in general different in the spin and charge sectors of
the theory.  Obviously
the group of conformal transformations in 2 + 1 dimensions is finite.
Consequently, the model does not possess the infinite symmetries of a true
1 + 1 dimensional conformal field theory.  But it is the current algebra
that concerns us most here.  It is the essential ingredient that permits
us to map Fermions onto Bosons and vice versa at each of the Fermi
points.
Again either Abelian or non-Abelian Bosonization works.  In this case
we choose non-Abelian Bosonization\refto{Ludwig}
for the spin sector by introducing
the Wess-Zumino-Witten (WZW) field $g^\alpha_\beta$
(the charge sector is still described by
an Abelian Boson $\phi$).  Non-Abelian Bosonization is superior in the
sense that it explicitly exhibits global SU(n) invariance.  With it we can
readily classify all SU(n) invariant operators.
(Abelian Bosonization hides SU(n) invariance
because an explicit choice for the spin
quantization direction must be made.)

The Bosonization dictionary again translates currents defined in terms
of the Fermi fields into Bosonic operators.  For example at
point 1 we have:
$$\eqalign{
J_1(u) &= \sqrt{{{2}\over{\pi}}}~ \partial_{u}~ \phi_1(u) \cr
J_{1 \alpha}^{\beta}(u) &= {{i}\over{4\pi}}~ [\partial_{u}~ g_{1
\gamma}^{\beta}
(u)] [g_{1 \alpha}^{\gamma}(u)] \ .\cr}
\eqno(2.10)$$
Here the free fixed point theory with all the $\lambda$'s equal to zero has
its charge sector described by a free Lagrangian density:
$${\it L}_c = {{1}\over{2L}}~ \{ (\partial_{u_+} \phi_1)^2
+ (\partial_{u_-} \phi_{-1})^2 + (\partial_{v_+} \phi_2)^2
+ (\partial_{v_-} \phi_{-2})^2 \}\ . \eqno(2.11a)$$
The spin sector consists of a
k = 1 WZW action given by $S_s[g_1, g_2, g_{-1}, g_{-2}]
\equiv S_{1}[g_1 + g_{-1}]
+ S_{2}[g_{2} + g_{-2}]$ where:
$$\eqalign{
S_{1}[g] &= {{1}\over{8\pi}} \int_{\partial V} du~ dt~ Tr~ \{
\partial_{u} g~ \partial_{u} g^\dagger + \partial_{t} g~
\partial_{t} g^\dagger \} \cr
&+ {{1}\over{12\pi}} \int_V du~ dt~ dz~ \epsilon^{\mu \nu \lambda}
Tr(g^\dagger \partial_\mu g~ g^\dagger \partial_\nu g~ g^\dagger
\partial_\lambda g) . \cr} \eqno(2.11b)$$
Here the second integral, the topological Wess-Zumino term,
is defined by extending
the domain of the g-field from physical two-dimensional (u, t)
space-time to a three dimensional volume V with space-time coordinates
$x_\mu = (u, t, z)$.  The boundary $\partial V$ of $V$ is taken to be the
(u, t) space-time.  Of course $S_{2}[g]$ is similar in form
to $S_{1}[g]$ but the spatial variable $v$ replaces $u$.
The spin sector of the free theory displays
$SU(n)_1 \otimes SU(n)_2 \otimes SU(n)_{-1} \otimes SU(n)_{-2}$ invariance
because the spin currents at the four points are decoupled.

Now the residual fixed line interactions $\lambda_{1c}$, $\lambda_{2c}$
and $\lambda_{2s}$ can be included by using the Bosonization rules of
Eq. [2.10].  In the Boson language, the 4-fold U(1) symmetry operation
amounts to a shift in each of the charged Boson fields by a constant:
$\phi_1(u, t) \rightarrow \phi_1(u, t) + \Gamma_1$, etc.
Since only derivatives of the Boson field appear in the action, it
continues to manifest four-fold U(1) invariance as expected.
On the other hand, local
$SU(n)_1 \otimes SU(n)_2 \otimes SU(n)_{-1} \otimes SU(n)_{-2}$ invariance
is at least partly broken by non-zero $\lambda_{2s}$
which couples together the zero-momentum components of the
spin currents at the four points.

Two issues remain to be investigated in our model:
First, how are Fermi statistics maintained in the Bosonization scheme,
now that there are two spatial directions?  And second, how do the residual
fixed point interactions change the character
of the Fermi points?  We answer these questions by constructing
a more general framework in the next section.

\vfill\eject
\centerline{III. LUTHER-HALDANE BOSONIZATION: INFINITE U(1) SYMMETRY}
\smallskip
Encouraged by the renormalization group flows in our model, we now
take a leap of faith, advocated most recently by Haldane\refto{Haldane0},
and postulate the existence of a similar fixed point, not
just for the four Fermi points, but rather for a continuum of Fermi points,
in other words, a Fermi surface.  We outline the construction of the
currents and the Hamiltonian first and later [in sections (IV) and (V)]
demonstrate that the framework reproduces well known results.

To be definite, we study the case of three spatial dimensions;
generalizations to other dimensions are straightforward.
We begin with the charge sector and study a smooth Fermi surface
parameterized by radial vectors $\bf S$
and $\bf T$ that label a fine, locally flat (and rectangular)
mesh of grid points on the Fermi surface with spacing
$\Lambda << k_f$ between the points.
We also place the system in a cubic box with sides of length $L$
and use periodic boundary
conditions so that
the momenta are quantized as ${\bf p_m} = {{2 \pi}\over{L}}~ {\bf m}$ where
$\bf m$ is a vector with integer components.
The most general charge Hamiltonian possessing infinite U(1) symmetry may then
be written as:
$$H_c = \half~ \sum_{\bf S,T} \sum_{\bf q} V_c({\bf S, T; q})~ J({\bf S; q})~
J({\bf T; -q}) .\eqno(3.1)$$
The prefactor of $\half$ compensates for over-counting due to the symmetry
of the summand under $\bf q \rightarrow -q$ with $\bf S \leftrightarrow T$.
The function $V_c({\bf S, T; q})$
encapsulates not only the Fermi velocity $v_f$ of the
non-interacting system but also the residual Fermi-liquid type
interactions $V'_c$ between quasi-particles.
The current at each point $J({\bf S; q})$ is now defined
in momentum space as:
$$J({\bf S; q}) \equiv \sum_{\bf k} \theta({\bf S; k + q})~
\theta({\bf S; k})~ \{ \psi^{\dagger \alpha}_{\bf k + q}~
\psi_{\alpha {\bf k}} - \delta^3_{\bf q, 0}~ n_{\bf k} \}
\ .\eqno(3.2)$$
The subtraction of the vacuum charge expectation value
$n_{\bf k} \equiv < \psi^{\dagger \beta}_{\bf k}
\psi_{\beta {\bf k}} >$ in Eq. [3.2] amounts to normal
ordering.  Our geometric construction of the currents involves tiling the
Fermi surface with squat rectangular pill boxes at each grid point $\bf S$.
The boxes have dimensions $\Lambda \times \Lambda$ along
the surface and extend in height $\pm \lambda/2$ above and below the
Fermi surface (see Figure [6]).  The function $\theta({\bf S; k}) = 1$ if
$\bf k$ lies inside the box; otherwise it is zero.  Thus $v_f \lambda$
functions
as an ultraviolet energy cutoff.

The current commutation relations may now be found by direct computation
with the use of the canonical anticommutation relation
$\{~ \psi^{\dagger \alpha}_{\bf q}~ ,~ \psi_{\beta \bf p}~ \} =
\delta^\alpha_\beta~ \delta^3_{\bf q, p}$.  First note that currents in
different patches commute because the Fermion operators that make
up the currents are also located in different patches. So:
$$\eqalign{[J({\bf S; q})~ ,~ J({\bf T; p})]
&= \delta^2_{\bf S,T}~ {\Bigl \lbrace} \sum_{\bf k} \theta({\bf S; k + q + p})
\theta({\bf S; k}) [\theta({\bf S; k + q}) - \theta({\bf S; k + p})]
\delta^3_{\bf q + p, 0} n_{\bf k} \cr
&\ \ \ \ + \sum_{\bf k} \theta({\bf S; k + q + p}) \theta({\bf S; k})
[\theta({\bf S; k + q}) - \theta({\bf S; k + p})] \cr
&\ \ \ \ \times (\psi^{\dagger \alpha}_{\bf k + q + p} \psi_{{\bf k} \alpha} -
\delta^3_{\bf q + p, 0} n_{\bf k}) {\Bigr \rbrace}
\ . \cr }\eqno(3.3)$$
In one dimension, the index $\bf S$ just labels the left and
right Fermi points and
the first sum in Eq. [3.3] is the usual quantum anomaly.  The second
sum vanishes in the limit of infinite bandwidth ($\lambda \rightarrow
\infty$) because in that case $\theta(q) = 1$ except very deep inside
or outside the Fermi sea.  In this limit, matrix elements of the
operator $(\psi^{\dagger \alpha}_{k + q + p} \psi_{k \alpha} -
\delta_{q + p, 0} n_k)$ vanish and we recover the usual one-dimensional
Kac-Moody algebra.  Thus, for the right movers,
$$[J(R; q)~ ,~ J(R; p)] = 2~ {{q~ L}\over{2 \pi}}~ \delta_{q+p,0}\eqno(3.4)$$
where the prefactor of 2 comes from the two spins.  We recognize this algebra
as the momentum space version of Eq. [1.7].

One might expect that the natural generalization of the currents to
two or three spatial dimensions would take the fields to be organized
along narrow rays of vanishing thickness radiating outward from the
center of the Fermi sea.  In fact this approach was adopted by Luther in
his pioneering work on the Bosonization of free Fermions
in higher dimension\refto{Luther} since
it reduces the higher dimensional problem to a set of simple
decoupled 1 + 1 dimensional systems.
However it is clear that the procedure breaks down when interactions of
the Fermi liquid type are included.
The charge Hamiltonian, Eq. [3.1], couples charge currents in
different boxes at positions $\bf S$ and $\bf T$.  As the Fermi surface
must have non-zero curvature, any wavevector $\bf q$ that lies inside
a tube at position $\bf S$, no matter how small, will be accompanied by
a wavevector $\bf -q$ that in general does {\it not} fit inside the tube
at a different point $\bf T$.

The problem is avoided with the use of the squat boxes.  The price we
pay for this new geometrical construction is the introduction of several limits
which must be carefully taken in order to arrive at the
correct commutation relations.  This delicate series of limits
in fact correspond to the Fermi liquid theory limits of $\omega \rightarrow
0$ and $|{\bf q}| \rightarrow 0$ such that ${{v_f |{\bf q}|}\over{\omega}}
\rightarrow 0$,
the so-called $\omega$-limit which pertains to collective modes rather
than quasi-particle scattering\refto{Baym}.
First we require the wavevectors $\bf q$ and $\bf p$ in Eq. [3.3] to be small:
$|{\bf q}| < {{\Lambda}\over{N}}$ and $|{\bf p}| < {{\Lambda}\over{N}}$
where we take $N \rightarrow \infty$.  Thus we may think of $\bf q$ as lying
within a small sphere inside the squat box -- see Figure [6].  The
limit insures that only the component of $\bf q$ normal to the surface
appears in the quantum anomaly.  As we shall see, it is the normal component
that is needed to
reproduce the spectrum of low-lying excitations in the free Fermion problem.
This geometrical result may be obtained by using the fact that
$n_{\bf k} = 2$ for
momenta $\bf k$ lying deep inside the Fermi sea and zero far outside.  With the
limits $|{\bf q}| < {{\Lambda}\over{N}}$ and $|{\bf p}| < {{\Lambda}\over{N}}$
we have
$\theta({\bf S; k+q+p}) \approx \theta({\bf S; k})$.  The sum
in the first term of Eq. [3.3] can be done and we have:
$$[J({\bf S; q})~ ,~ J({\bf T; p})] = 2~ \delta_{\bf S, T}~ [~
\delta^3_{\bf q + p, 0}~
\Lambda^2~ ({{L}\over{2 \pi}})^3~ {\bf q \cdot \hat{n}_S}~ +~ error~
term~ ] . \eqno(3.5a)$$
Here $\bf \hat{n}_S$ is the normal vector pointing outward at
point $\bf S$ on the
Fermi surface and the error term is the second sum in Eq. [3.3]
which ruins the Kac-Moody algebra
because it is not a c-number but rather an operator involving the Fermi fields
$\psi^\dagger$ and $\psi$.
Note that with the above limit on the size of $\bf q$ the magnitude of the
quantum anomaly is of order ${{\Lambda^3}\over{N}}~ ({{L}\over{2 \pi}})^3$.

Let us estimate now the size of the error term in the commutation relations.
It may be estimated by replacing
the operator $(\psi^{\dagger \alpha}_{\bf k + q + p} \psi_{{\bf k} \alpha} -
\delta^3_{\bf q + p, 0} n_{\bf k})$ by $1 - \delta^3_{\bf q + p, 0}$
and computing the volume of the geometrical complement of the intersection
of two $\theta$ functions $[\theta({\bf S; k + q}) - \theta({\bf S; k + p})]$
appearing in the second sum.  Note that the tops and bottoms of the boxes
do not contribute because the matrix elements are assumed to be zero
deep inside or outside the Fermi sea.  Only the sides of the pill boxes matter.
A simple computation then shows that this term is off order
${\rm Max}\{|{\bf q}|, |{\bf p}| \}~\times~
\Lambda~ \lambda~ ({{L}\over{2 \pi}})^3 <
{{\Lambda^2~ \lambda}\over{N}}~ ({{L}\over{2 \pi}})^3$.  Therefore, the choice
of a squat pill box with $\Lambda = \sqrt{N} \lambda$ makes the error term
small (of order ${{1}\over{\sqrt{N}}}$) in comparison to the quantum anomaly.
This second limit is equivalent to the ``$\omega$-limit'' of Fermi liquid
theory as ${{\omega}\over{v_f}}$ corresponds to $\lambda$ which is now of order
$\sqrt{N}$ times larger than the momentum $\bf q$.
It is satisfying to have this simple geometrical interpretation of the
$\omega$-limit of Fermi liquid theory.

The current algebra Eq. [3.5a] can be put into a more familiar form with a
Fourier transform over the two components of the momentum perpendicular to the
Fermi surface normal vector, $\bf q_\perp$ and $\bf p_\perp$.  Then we obtain:
$$[J({\bf S}; q_\parallel, {\bf x_\perp})~ ,~ J({\bf T};
p_\parallel, {\bf y_\perp})]
= 2~ \delta_{\bf S, T}~ {{q_\parallel L}\over{2 \pi}}~
\delta_{q_\parallel + p_\parallel, 0}~
L^2 \delta^2({\bf x_\perp - y_\perp})\ . \eqno(3.5b)$$
Here the current
algebra is identical to the usual one-dimensional one, Eq. [3.4], except with
additional labels $\bf S$, $\bf T$,
$\bf x_\parallel$ and $\bf y_\parallel$ that ``come along
for the ride.'' Thus the well-developed
theory of one-dimensional Kac-Moody algebra representations\refto{GO} applies
equally well to our generalized algebra and we can use this machinery
to find the spectrum of states.  Of course state counting is
simple in the abelian case but null states appear in representations of
the non-abelian Kac-Moody algebra.

What choice of parameters $V_c$ yield the correct spectrum for the charge
sector?
The non-interacting limit is recovered by making the following choice:
$V_c({\bf S, T; q}) = \half v_f({\bf S})~ \Omega^{-1}~ \delta^2_{\bf S, T}.$
Here the factor of $\Omega \equiv \Lambda^2 ({{L}\over{2 \pi}})^3$
cancels the factors of volume appearing in
the current algebra Eq. [3.5] and the factor of $\half$ compensates for the
2 due to up and down spins.
With the algebra Eq. [3.5] we then recover the free dispersion relation
$\omega({\bf S; q}) = v_f({\bf S})~ {\bf q \cdot \hat{n}_S} \equiv {\bf v_S
\cdot q}.$  To scale the interaction coefficients properly,
we appeal to Fermi liquid theory and note that the current evaluated at zero
momentum is equal to the occupancy fluctuation operator
summed over the interior of the pill box:
$$J({\bf S; q = 0}) = \sum_{\bf k}~ \theta({\bf S; k})~ \delta n_{\bf k}
\eqno(3.6)$$
where $\delta n_{\bf k} \equiv
\psi^{\dagger \alpha}_{\bf k} \psi_{\alpha {\bf k}} -
< \psi^{\dagger \alpha}_{\bf k} \psi_{\alpha {\bf k}} > $.
Therefore, the Fermi liquid interaction
is identical to the zero-momentum piece of our interaction term:
$${{1}\over{2 L^3}}~ \sum_{\bf k, p}~ f({\bf k, p})~
\delta n_{\bf k}~ \delta n_{\bf p} =
{{1}\over{2 L^3}}~ \sum_{\bf S, T}~ V'_c({\bf S, T; q = 0})~ J({\bf S; 0})~
J({\bf T; 0}) \eqno(3.7)$$
if we identify $f({\bf k_S, p_T}) = V'_c({\bf S, T; 0})$ by assuming that
the Fermi liquid interactions depend only on the momenta $\bf k_S$ and
$\bf p_S$ at points $\bf S$ and $\bf T$ of the Fermi surface,
not the component of the momentum perpendicular to the surface.
Evidently, a factor of inverse volume (${{1}\over{L^3}}$) should be included
in the interaction term,
$$V_c({\bf S, T; q}) = \half v_f({\bf S})~ \Omega^{-1}~ \delta^2_{\bf S, T}
+ {{1}\over{L^3}}~ V'_c({\bf S, T; q})\ .\eqno(3.8)$$
Note that while our theory contains the same Fermi surface interactions as
traditional Landau Fermi liquids, the form of the interaction is more
general than Fermi liquid theory as it depends on ${\bf q}$,
the momentum of the collective excitation.  In Fermi liquid theory,
the parameters
$f({\bf k, p})$ do not depend on $\bf q$ and the  momenta
$\bf k$ and $\bf p$ appearing in $f({\bf k, p})$ are constrained
to lie on the Fermi surface.  A {\it different} extension of Fermi
liquid theory which relaxes this constraint on $\bf k$ and $\bf p$
is described in the next section.
Our calculation of non-analytic contributions to the specific heat will
highlight the difference between these two types of generalizations.

We see therefore that in higher dimension, as in one dimension, the
Bosonization procedure puts the free and interacting
components of the Hamiltonian on an equal footing, despite the fact that
the free piece is quadratic in the Fermion operators while the interaction
is quartic.  This simplicity is a result of the current algebra Eq. [3.5]
which permits us to express both terms as bilinears in the currents.
It is however somewhat deceptive because more general quartic
terms, for instance the $\lambda_5$ interaction in our simplified
model of the preceding section,
cannot be expressed as bilinears in the current operators.  Nevertheless,
these interactions have a Bosonic representation, albeit a more complicated
one.
We show how to Bosonize general interactions below.

First we focus on the spin sector.  The total Hamiltonian is a given by the
sum of the charge and spin Hamiltonians.  To form the spin Hamiltonian, we
define spin currents.  In the general SU(n) case we have:
$$J^\alpha_\beta({\bf S; q}) \equiv \sum_{\bf k} \theta({\bf S; k + q})~
\theta({\bf S; k})~ \{ \psi^{\dagger \alpha}_{\bf k + q}~
\psi_{\beta {\bf k}} - {{1}\over{n}}~ \delta^\alpha_\beta~
\psi^{\dagger \gamma}_{\bf k + q} \psi_{\gamma {\bf k}} \}
\ .\eqno(3.9)$$
Like the charge currents, spin currents at different grid points commute,
but the non-Abelian Kac-Moody algebra governs currents at the same point
in the $N \rightarrow \infty$ limit:
$$\eqalign{[J^\alpha_\beta({\bf S; q})~ ,~ J^\gamma_\delta({\bf S; p})]
&= (\delta^\alpha_\delta~ \delta^\gamma_\beta~ - {{1}\over{n}}~
\delta^\alpha_\beta~ \delta^\gamma_\delta)
\delta^3_{\bf q + p, 0}~ \Omega~ {\bf q \cdot \hat{n}_S} \cr
&~ ~ ~ + \delta^\gamma_\beta J^\alpha_\delta({\bf S; q + p})
- \delta^\alpha_\delta J^\gamma_\beta({\bf S; q + p}) \cr}
\ .\eqno(3.10)$$
The physical SU(2) Kac-Moody algebra can be expressed more succinctly as:
$$[J^a({\bf S; q})~ ,~ J^b({\bf S; p})]
= \half~ \delta^{a b} \Omega~ {\bf q \cdot \hat{n}_S}
+ i \epsilon^{a b c}~ J^c({\bf S; q + p}) \ .\eqno(3.11)$$
The spin Hamiltonian may then be written:
$$H_s = \half~ \sum_{\bf S,T} \sum_{\bf q} V_s({\bf S, T; q})~
{\bf J}({\bf S; q})~ {\bf \cdot~ J}({\bf T; -q})\ \eqno(3.12)$$
where $V_s$ incorporates the Fermi velocity of spin excitations
and spin-spin interactions at different points on the Fermi surface:
$$V_s({\bf S, T; q}) = {{2}\over{3}}~
v_f({\bf S})~ \Omega^{-1}~ \delta^2_{\bf S, T}
+ {{1}\over{L^3}}~ V'_s({\bf S, T; q})\ .\eqno(3.13)$$
In general it is not possible to exactly diagonalize the Hamiltonian;
the non-Abelian nature of the algebra precludes this.  We encountered this
problem in a simpler form in section (II) where
$\lambda_{1s}$, the parameter that couples together
spin currents on opposing Fermi points, flows by itself (see Eq. [2.9]).

Both the charge and spin currents are invariant under the local U(1) operation
which changes
the phase of all the Fermions inside a given pill box by the same
(time-independent) amount $\Gamma$.  If $\bf k$ lies inside
the box centered at grid point $\bf S$
then $$\psi_{\alpha}({\bf k}) \rightarrow e^{i \Gamma({\bf S})}~
\psi_{\alpha}({\bf k})\ [{\rm only when}~ \theta({\bf S; k}) = 1]\eqno(3.14)$$
leaves the currents invariant because the $\psi^{\dagger}$
fields, which transform with the opposite phase factor, cancel the overall
phase change.
Thus the Hamiltonian is automatically invariant under the infinite
U(1) symmetry.  The physical meaning of the invariance is clear: the Fermi
liquid type interactions preserve the Fermion occupancy at each point in
momentum space because quasi-particle scattering is suppressed in the
$N \rightarrow \infty$ $\omega$-limit.  The U(1) symmetry just reflects the
local conservation of particle number.

Indeed, it is the existence of an
infinite number of conservation laws that makes the charge
sector of the problem solvable.  On the other hand, the free Fermion
system also exhibits local SU(2) symmetry.  So it is rather surprising to
discover that the spin current interactions in general break the infinite local
SU(2) invariances down to a single global SU(2) symmetry.  The local
invariance is broken because spin currents at different points on the Fermi
surface must rotate together to keep the spin Hamiltonian Eq. [3.12] invariant.
The special case of purely local current-current coupling,
$V_s({\bf S, T; q}) = {{2}\over{3}}~
v_f({\bf S})~ \Omega^{-1}~ \delta^2_{\bf S, T}$,
is an exception which restores the full local
SU(2) invariance.  As expected, the Hamiltonian is
now exactly solvable: the Hamiltonian describes
free spin excitations propagating at the Fermi velocity.
For this special case only the quantum
anomaly in Eq. [3.11], $\half \delta^{ab} \Omega {\bf q \cdot \hat{n}_S}$,
not the $i \epsilon^{abc}~ J^c({\bf S; q+p})$ term, is important
because of the symmetry $\bf q \rightarrow -q$.
The factor ${{2}\over{3}}$ compensates for the three spin components.
The spectrum of states may now be found either by simply choosing a
spin-quantization axis, or in an SU(2) invariant manner with the use of
Kac-Moody representation theory.\refto{GO}

Actually, we can find the excitation spectrum when the interactions
described by $V'_s$ are non-singular.  In this case, we may treat the
spin currents as semi-classical objects: the right-hand side of the
commutator Eq. [3.11] can be set to zero by rescaling the currents to
be of order one.  The problem resembles the large-spin limit of a
quantum magnet since the currents incorporate a sum over
$\lambda \Omega >> 1$ points in momentum space.  If we rescale
$J^a({\bf S; q}) \rightarrow (\lambda \Omega)^{-1}~
J^a({\bf S; q})$ then the rescaled currents obey:
$$\eqalign{[J^a({\bf S; q})~ ,~ J^b({\bf S; p})]
&= {{1}\over{(\lambda \Omega)^2}}~ \{~
\half \delta^{a b} \Omega~ {\bf q \cdot \hat{n}_S}
+ i \epsilon^{a b c}~ J^c({\bf S; q + p}) \}\cr
&\rightarrow 0\cr}\eqno(3.15)$$
as $L \rightarrow \infty$ with $\Lambda$ held fixed.
The emergence of the classical limit should
not be surprising; after all, Landau Fermi Liquid Theory is essentially
classical in nature.  The free dispersion is still determined by the quantum
anomaly; only the interactions are treated classically by replacing
the current operators with their expectation values $J^a({\bf S; q})
\rightarrow \langle J^a({\bf S; q}) \rangle$ evaluated in the excited
state of interest.
This procedure trivially reproduces the excitation spectrum of
Fermi Liquid Theory.  Note, however, that the classical limit breaks down
in the case of singular interactions.  For example, the first term in
Eq. [3.13],
${{2}\over{3}}~ v_f({\bf S})~ \Omega^{-1}~ \delta^2_{\bf S,T}$,
is singular because the factor
$\Omega^{-1}$ diverges as the number of mesh points increases.
In this case the quantum anomaly cannot
be neglected and in fact is needed to reproduce the free dispersion relation.
Likewise, any singular spin current interactions that couple
different patches on the Fermi surface destroy the classical limit:
the small anomaly cannot be neglected because of the
large interaction.  In fact this is the
generic situation in one spatial dimension, where interactions that couple
the left and right points are generally of the same order as current-current
terms that involve
only one point.  In other words, there is no sense in which the
Fermi-Liquid type interactions can be smooth when there are just two
Fermi points.  We return to this point in the discussion of section (VI).

By introducing Boson fields conjugate to the currents,
the Fermi fields and interaction terms
can be Bosonized\refto{Luther,Haldane0}.  We proceed by
analogy to our construction in one-dimension [section (I)] and concentrate
on spinless Fermions; it is straightforward to include spin
via either Abelian or non-Abelian Bosonization.
We introduce the coarse-grained
Boson field $\phi({\bf S; x})$ and the associated
Boson current in the direction normal to the Fermi surface:
$$J({\bf S; x}) = \sqrt{4 \pi}~
{\bf \hat{n}_S \cdot \nabla} \phi({\bf S; x})\ .\eqno(3.16)$$
The Boson field is related to
the microscopic fields $\phi({\bf p})$ by
coarse graining over the pill box:
$$\phi({\bf S; x}) = {{\sqrt{\Omega}}\over{2\pi}}~ \sum_{\bf p, p \cdot
\hat{n}_S > 0}~
{{\theta({\bf S; p})}\over{2 \sqrt{|{\bf p \cdot \hat{n}_S}|}}}~
\{e^{i {\bf p \cdot x}} \phi({\bf p}) +
e^{-i {\bf p \cdot x}} \phi({\bf -p})\}\ .\eqno(3.17)$$
The microscopic Boson fields satisfy equal-time commutation relations:
$$[\phi({\bf p})~ ,~ \phi({\bf q})]
= \delta^3_{\bf p+q,0}\ .\eqno(3.18a)$$
Note that the reality of the microscopic fields $\phi({\bf x})$ means
that $\phi({\bf -k}) = \phi^\dagger({\bf k})$ and with this in mind the
commutation relations Eq. [3.18a] take on the more familiar form:
$$[\phi({\bf p})~ ,~ \phi^\dagger({\bf q})]
= \delta^3_{\bf p,q}\ .\eqno(3.18b)$$
Consequently, the coarse-grained fields obey a natural three-dimensional
generalization of the one-dimensional equal-time commutation relations
Eq. [1.8]:
$$[\phi({\bf S; x})~ ,~ \phi({\bf T; y})] = i {{\Omega}\over{4}}~
{{L^3}\over{2 \pi}}~ \delta^2_{\bf S,T}~
\epsilon({\bf \hat{n}_S \cdot [x - y]})~
\delta^2({\bf x_\perp - y_\perp})\eqno(3.19a)$$
where again $\epsilon(x) = 1$ for $x > 0$; otherwise it equals $-1$.
Here $\bf x_\perp$ denotes the two components of $\bf x$ that are
perpendicular to the surface normal $\bf \hat{n}_S$.  Note that
$\delta^2(0) = ({{\Lambda}\over{2 \pi}})^2$ which is the area of the base
of the pill box.  Thus, when $\bf x_\perp = y_\perp$ we have:
$$[\phi({\bf S; x})~ ,~ \phi({\bf T; y})] = i {{\Omega^2}\over{4}}~
\delta^2_{\bf S,T}~ \epsilon({\bf \hat{n}_S \cdot [x - y]})~ ;\eqno(3.19b)$$
otherwise the $\phi({\bf S; x})$ fields commute.
Furthermore, the Boson currents Eq. [3.16] satisfy the same U(1) Kac-Moody
algebra Eq. [3.5]
as the Fermion charge currents Eq. [3.2] (with half the anomaly because
we have removed the spin index):
$$[J({\bf S; q})~ ,~ J({\bf T; p})] = \Omega~ \delta_{\bf S, T}~
\delta^3_{\bf q + p, 0}~ {\bf q \cdot \hat{n}_S}\eqno(3.20a)$$
or in real space,
$$[J({\bf S; x})~ ,~ J({\bf T; y})] = -i~ \Omega~ \delta_{\bf S, T}~
L^3 {\bf \hat{n}_S \cdot \nabla_x}~ \delta^3({\bf x - y})\ .\eqno(3.20b)$$
Here the Fourier transform of the currents is given by:
$$J({\bf S; x}) \equiv
\sum_{\bf q}~ e^{i {\bf q \cdot x}}~ J({\bf S; q})\ .\eqno(3.21)$$
The Hamiltonian Eq. [3.1] then becomes (for spinless Fermions):
$$H = 2 \pi \int~ d^3x~ d^3y\sum_{\bf S,T} V_c({\bf S, T; x-y})~
[{\bf \hat{n}_S \cdot \nabla} \phi({\bf S; x})]~
[{\bf \hat{n}_T \cdot \nabla} \phi({\bf T; y})]\eqno(3.22)$$
and Fermi fields are expressed in terms of the Boson fields as:
$$\psi({\bf S; x}) = {{\Omega^\half}\over{\sqrt{2 \pi a}}}~
e^{i {\bf k_S \cdot x}}~ {\rm exp}\{i {{\sqrt{4 \pi}}\over{\Omega}}~
\phi({\bf S; x})\}
\eqno(3.23)$$
where $\bf k_S$ is the Fermi momentum at grid point $\bf S$.

The N-point Fermion correlation functions are reproduced with the use of
the Bosonization formula Eq. [3.23].  If, for example,
we use the operator identity:
$$e^A~ e^B = : e^{A + B} :~ {\rm exp}
\langle A B + \half (A^2 + B^2) \rangle \eqno(3.24)$$
then we find that the Fermion two-point function is given by:
$$\langle \psi^\dagger({\bf S; x})~ \psi({\bf T; 0}) \rangle
= {{1}\over{2 \pi a}}~ \delta^2_{\bf S,T}~ e^{i {\bf k_S \cdot x}}~
{\rm exp} \{ {{4 \pi}\over{\Omega^2}}~ \langle \phi({\bf S; x})
\phi({\bf S; 0}) - \phi^2({\bf S; 0}) \rangle \}\ .\eqno(3.25)$$
The Boson correlation function can be computed using the relation
Eq. [3.17] and the result is:
$$\eqalign{G_c({\bf S; z})
&\equiv \langle \phi({\bf S; x})~ \phi({\bf S; 0})
- \phi^2({\bf S; 0})\rangle\cr
&= -{{\Omega^2}\over{4 \pi}}~ {\rm ln}({{{\bf \hat{n}_S \cdot x} + i a}
\over{ia}})\ ;\ |{\bf x_\perp} \Lambda| << 1 \cr
&\rightarrow -\infty\ ;\ |{\bf x_\perp} \Lambda| >> 1\ .\cr}\eqno(3.26)$$
Consequently we obtain the correct Fermion correlation function,
coarse-grained over the pill box:
$$\langle \psi^\dagger({\bf S; x})~ \psi({\bf T; 0}) \rangle
= {{i \Omega}\over{2 \pi}}~ \delta^2_{\bf S,T}~
{{e^{i {\bf k_S \cdot x}}} \over{{\bf k_S \cdot x} + i a}}~
({{2 \pi}\over{\Lambda}})^2~ \delta^2({\bf x_\perp})\ .\eqno(3.27)$$
It should be emphasized that it is the average over the pill box that
results in the $\delta^2({\bf x_\perp})$ term.

To close the circle (Bosons $\rightarrow$ Fermions $\rightarrow$ Bosons)
we form the Fermion current Eq. [3.2].  In
real space we utilize the point-splitting procedure:
$$\eqalign{J({\bf S; x})
&= :\psi^\dagger({\bf S; x}) \psi({\bf S; x}):\cr
&= {{\Omega}\over{2 \pi a}}~ {\rm lim}_{\epsilon \rightarrow 0}~
:{\rm exp}[-i {{\sqrt{4 \pi}}\over{\Omega}}~
\phi({\bf S; x + \hat{n}_S}\epsilon)]~
{\rm exp}[i {{\sqrt{4 \pi}}\over{\Omega}}~ \phi({\bf S; x})]:
\cr}\eqno(3.28)$$
then using the operator identity Eq. [3.24] again we obtain:
$$\eqalign{J({\bf S; x})
&= {{\Omega}\over{2 \pi a}}~ {\rm lim}_{\epsilon \rightarrow 0}~
{\rm exp}[-i {{\sqrt{4 \pi}}\over{\Omega}}~ \{\phi({\bf S; x
+ \hat{n}_S}\epsilon) - \phi({\bf S; x})\}]~
{\rm exp}[{{4 \pi}\over{\Omega^2}}~ G_c({\bf S; x})] \cr
&= {{1}\over{\sqrt{\pi}}}~ {\bf \hat{n}_S \cdot \nabla} \phi({\bf S; x})
\cr}\eqno(3.29)$$
which is identical to Eq. [3.16].  A similar calculation shows that the
free Fermion Hamiltonian is of the same form as
the Boson Hamiltonian Eq. [3.22]:
$$\eqalign{H^0
&\equiv \sum_{\bf S}~ v_f({\bf S})~ \int d^3x~ \psi^\dagger({\bf S; x})
({\bf \hat{n}_S \cdot \nabla}) \psi({\bf S; x}) \cr
&= 2 \pi \sum_{\bf S}~ {{v_f({\bf S})}\over{\Omega}}~ \int d^3x~
\{ ({\bf \hat{n}_S \cdot \nabla}) \phi({\bf S; x})\}^2\ .\cr}\eqno(3.30)$$

As it stands, $\psi$ fields located in the same patch and at the
same perpendicular coordinates anticommute.  For example,
the Fermion two-point function Eq. [3.27] is odd under the transformation
$\bf x \rightarrow -x$ followed by complex conjugation which is equivalent
to interchanging the creation and annihilation operators in a translationally
invariant system.
However, fields in different patches, and fields in the same patch
with $\bf x_\perp \neq y_\perp$, {\it commute}:
$$\eqalign{\{\psi({\bf S; x})~ ,~ \psi({\bf S; y})\} = 0~ ;~
{\bf x_\perp = y_\perp}\cr
[\psi({\bf S; x})~ ,~ \psi({\bf S; y})] = 0~ ;~ {\bf x_\perp \neq y_\perp}\cr
[\psi({\bf S; x})~ ,~ \psi({\bf T; y})] = 0;~ ~ {\bf S} \neq {\bf T}\ .\cr}
\eqno(3.31)$$
The commutation relations can be transformed into the correct anticommutation
relations by introducing
an ordering operator analogous to a Jordan-Wigner transformation\refto{Luther}.
Let $O({\bf S})$ be the ordering operator defined by:
$$O({\bf S}) \equiv {\rm exp} \{i~ {{\pi}\over{2}}~
\sum_{\bf T = 1}^{\bf S - 1}~
J({\bf T; q = 0}) \} \eqno(3.32)$$
where the mesh points $\bf T$ have been arranged in consecutive order.
To be definite, we could follow Luther's prescription
and choose the mesh points to begin at
some point (the ``north pole'')
on the Fermi surface, spiral outwards, and converge at the
antipode (``south pole'').
It is straightforward to check that the combination
$\psi({\bf S; x}) O({\bf S})$ anticommutes with $\psi({\bf T; y}) O({\bf T})$
when $\bf S \neq T$.  Commuting statistics are still obeyed when
the fields are in the same pill box, but this discrepancy can be neglected
in the continuum limit $\Lambda \rightarrow 0$.  Alternatively, a second
ordering operator may be introduced to implement anticommuting statistics
within the pill box.

Thus we see that charge sector of the semi-classical Landau theory
has been replaced by a quantum mechanical theory.  The Fermi
liquid should be thought of as a zero-temperature quantum critical
Gaussian fixed point
with infinite U(1) symmetry and parameters $V_c({\bf S, T; q})$.
No longer do semi-classical entities like $\delta n_k$ appear:
these have been replaced
by charge current operators that are quantized with the Kac-Moody algebras.
On the other hand, we have to resort to a semi-classical description of
the spin sector because the quantum version appears to be intractable.
A geometrical meaning has been given to the $\omega$-limit and a direct
connection between the quasiparticle operators and the Boson fields is made
via the Bosonization formulas.  To exercise the new framework, we rederive
some well known results in the next two sections.  We concentrate on the
charge sector to illustrate how the quantum theory reproduces these results.

\vfill\eject
\centerline{IV. $T^3 {\rm ln}(T)$ CONTRIBUTION TO THE SPECIFIC HEAT}
\smallskip
As a concrete application of the proceeding formalism, we calculate
the specific heat of an interacting Fermi liquid in three
spatial dimensions.
We obtain a non-analytic $T^3 {\rm ln}(T)$ contribution to the specific heat.
The existence of such a term is consistent with careful
measurements\refto{Greywall} of the specific heat in Helium-3.

We turn off the spin-spin interactions in the following and for simplicity
eliminate the spin index.
As the nonanalytic behavior arises from small momentum processes, it is
permissible to treat the Fermi surface in a locally flat approximation.
Let the surface normal point in the $\hat{z}$ direction.
Then the U(1) Kac-Moody algebra can be written as:
$[J({\bf S; q}),~ J({\bf T; p})] = 2 \Omega~ q_z~ \delta^2_{\bf S, T}~
\delta^3_{\bf p + q, 0}$.  These commutation relations are equivalent to those
obeyed by Bosonic
harmonic oscillator creation and annihilation operators once we
rescale by a factor of the square root of the momentum perpendicular to
the surface:
$$\eqalign{
J({\bf S; q}) &= \sqrt{-2 \Omega~ q_z}~
a^\dagger({\bf S; -q})~ ;\ \ q_z \leq 0 \cr
&= \sqrt{2 \Omega~ q_z}~ a({\bf S; q})~ ; \ \ q_z > 0 \cr}
\eqno(4.1)$$
where $[a({\bf S; q}),~ a^\dagger({\bf T; p})] = \delta^2_{\bf S, T}~
\delta^3_{\bf p, q}$.
Thus we can find the spectrum by direct diagonalization of the Bosonic harmonic
oscillator Hamiltonian:
$$H_c = \sum_{\bf S, T}~ \sum_{{\bf q},q_z > 0}~ V_c({\bf S, T; q})~
(2 \Omega)~ q_z~ a^\dagger( {\bf S; q})~ a({\bf T; q})\ . \eqno(4.2)$$
We again place the system in a box of dimensions $L^3$ and use
periodic boundary
conditions so the momenta are quantized as ${\bf q_m} = {{2 \pi}\over{L}}~
{\bf m}$.
The Fermi velocity is given in terms of the Fermi energy
$\epsilon_f$ by $v_f = \sqrt{2 \epsilon_f / m}$
and the number of states at the Fermi surface,
$A$, is given by $A \equiv
\sum_{\bf S} \sum_{\bf q}~ \theta({\bf S; q})~ \delta_{\bf \hat{n}_S
\cdot q , 0}
= {{2 m \epsilon_f}\over{\pi}}~ L^2
= {{p_f^2}\over{\pi}}~ L^2$.
Because the pill boxes completely tile the surface we have the sum rule:
$$\sum_{\bf S}~ \sum_{\bf q} \theta({\bf S; q})
= A~ {{\lambda L}\over{2 \pi}} \ .\eqno(4.3)$$
The specific heat is computed by using the
standard formula for {\it Bosons}:
$$C_V = {{1}\over{4 k_B T^2}}~ \sum_{\bf S}~ \sum_{{\bf q},q_z > 0}~
\theta({\bf S; q})~
{{\epsilon^2({\bf S; q})}\over{{\rm sinh}^2({{\epsilon({\bf S; q})}
\over{2 k_B T}})}}
\eqno(4.4)$$
where $\epsilon({\bf S; q})$ is an eigenvalue of the Hamiltonian Eq. [4.2]
which depends on the
momentum $\bf q$ as well as the index $\bf S$ that labels the vector
space of the patches covering the Fermi surface.
Let us first consider the case of non-interacting Fermions.
The eigenenergies of the Bosonized Hamiltonian are simply:
$$\epsilon({\bf S; q}) = v_f~ {\bf q \cdot \hat{n}_S}\ .
\eqno(4.5)$$
The sum over the patch index $\bf S$ and the components of $\bf q$ parallel
to the surface just yields the number of states at the Fermi surface, $A$.
The sum over the component of $\bf q$ perpendicular to the surface can be
converted to an integral.  Assuming that the temperature is small (so that
the thermally excited particle-hole pairs lie within the pill box, in other
words $k_B T << v_f \lambda$) we then
find:
$$\eqalign{C_V &= {{A}\over{4 k_B T^2}}~ {{L}\over{2 \pi}}~ \int_0^\infty
{{v_f^2 q_z^2}\over{{\rm sinh}^2({{v_f q_z}\over{2 k_B T}})}}~ dq_z \cr
&= {{A}\over{4 k_B T^2}}~
{{(2 k_B T)^3}\over{v_f}}~ {{L}\over{2 \pi}}~ (\pi^2/6) \cr
&= k_B^2 T~ ({{m p_f}\over{2 \pi^2}})~ {{\pi^2}\over{3}}~ L^3\
.\cr}\eqno(4.6)$$
This result is the correct answer for spinless Fermions and of course it should
be multiplied by a factor of two to account for the spin.  It is remarkable
that the Boson formula, Eq. [4.4], yields the full specific heat.  We
take it as further evidence that even for spatial dimensions greater than one
Bosonization reproduces the entire Fermion Hilbert space.

We now follow Pethick and Carneiro\refto{PC} and
focus on quasi-particles separated only by a small momentum
${\bf W} \equiv {\bf k_S - k_T} $ (ie. $|{\bf W}| << k_f$)
since a consideration of these processes is sufficient to demonstrate the
existence of non-analytic contributions to the specific heat.  Define two
vectors ${\bf u} \equiv {\bf k_S + q}$ and ${\bf u + p} \equiv {\bf k_T - q}$.
The quantity $\bf \hat{u} \cdot \hat{p}$ then functions as a small,
rotationally-invariant, dimensionless expansion
parameter.
Here the normalized momenta are defined by
${\bf \hat{p}} \equiv {\bf p} / p$ where
$p \equiv |{\bf p}|$.  Figure [7] exhibits the geometry of the interaction.
The interaction coefficient $V'_c$ may be expanded in our cylindrical
coordinate system.
Note that odd powers of $\bf \hat{u} \cdot \hat{p}$ do not
appear in the expansion because the sum over grid points and momentum
eliminates terms odd in $\bf p$.
$$\eqalign{V'_c({\bf S, T; q})
&= a + b~ ({\bf \hat{u} \cdot \hat{p}})^2 + ... \cr
&= a + b~ {{4 q_z^2}\over{4 q_z^2 + {\bf W}^2}} + ... \cr}
\eqno(4.7)$$
The expansion parameter is controlled in the low-temperature limit
which keeps $q_z$, the particle-hole momentum perpendicular to
the Fermi surface, small.  (Recall that squat pill-boxes force
$|{\bf W}| >> |{\bf q}| \geq |q_z|$ in the $N \rightarrow \infty$ limit.)

Actually, the interaction differs from the one that Pethick and Carneiro
studied: it couples particle-hole pairs at points $\bf S$ and $\bf T$ whereas
the Pethick-Carneiro interaction couples the occupancies
$n_{\bf u}$ and $n_{\bf u + p}$.  To be precise, the Pethick-Carneiro
interaction has the form:
$$\sum_{\bf u, p}~ b~ ({\bf \hat{u} \cdot \hat{p}})^2~ \delta n_{\bf u}~
\delta n_{\bf u + p}\ .\eqno(4.8)$$
This interaction cannot be directly expressed in terms of the currents since it
involves products of distinct occupancies above and below the Fermi surface
whereas
the current operator evaluated at zero momentum, $J({\bf S; 0})$, averages the
occupancy operator over the interior of the pill box.  Therefore a direct
connection with the earlier calculation cannot be made.  Nevertheless, our
purpose here is to show how non-analytic contributions to the
specific heat arise in the new framework.  Other terms may make non-analytic
contributions; the interaction Eq. [4.7] is the simplest such term within our
framework.

To proceed we diagonalize the Hamiltonian
with the aid of a Fourier transform from Fermi surface
patch index $\bf S$ space to $\bf X$-space.
Let $$a^\dagger({\bf S; q}) = {{\Lambda}\over{\sqrt{4 \pi p_f^2}}}~
\int d^2X~ e^{-i {\bf X \cdot S}}~ a^\dagger({\bf X; q})\ \eqno(4.9)$$
where $\int d^2X~ 1 = {{4 \pi p_f^2}\over{\Lambda^2}} = ({\rm number~ of~
patches})$ then
$$H = L^3~ \int d^2X~ \int_{-\Lambda/2}^{\Lambda/2} {{dq_x}\over{2\pi}}~
\int_{-\Lambda/2}^{\Lambda/2} {{dq_y}\over{2\pi}}~
\int_0^\lambda {{dq_z}\over{2\pi}}~
\epsilon({\bf X; q})~ a^\dagger({\bf X; q})~ a({\bf X; q})\ .\eqno(4.10)$$
Using Eq. [3.8] and Eq. [4.1] we then obtain the eigenenergies:
$$\epsilon({\bf X; q}) =
v_f q_z + {{8 b \Lambda^2 q_z^3}\over{(2 \pi)^3}}
\sum_{\bf W}~
{{e^{-i {\bf X \cdot W}/\Lambda}}\over{4q_z^2 + {\bf W}^2}}\ .\eqno(4.11)$$
The sum can be converted to a Riemann integral with the substitution
$\Lambda^2~ \sum_{\bf W}~ \rightarrow~ \int d^2 W$ and we find:
$$\epsilon({\bf X; q}) = v_f q_z - {{8 b q_z^3}\over{(2 \pi)^3}}~
\pi~ {\rm ln}(4 q_z^2~ {\bf X}^2)\ .\eqno(4.12)$$
In this equation we discard uninteresting terms proportional
to $b$ that make additional analytic contributions to the specific heat
and keep only the logarithmic piece.
We treat this term as a perturbation and calculate the specific heat
to O(b); then the change in the specific heat $\delta C_V$ due to the
perturbation is:
$$\eqalign{\delta C_V &\approx -b {{A}\over{4 k_B T^2}}~
{{32 v_f \pi}\over{(2\pi)^3}}~
{{L}\over{2\pi}}~ \int_0^\infty dq_z~ {{q_z^4~ {\rm ln}(q_z)}\over{
{\rm sinh}^2({{v_f q_z}\over{2 k_B T}})}} \cr
&\approx -b {{16 A L k_B^4}\over{\pi^3 v_f^4}}~ T^3 {\rm ln}(T)
\int_0^\infty~ {{x^4}\over{{\rm sinh}^2(x)}}~ dx \ .\cr}\eqno(4.13)$$
In the second line we retain only the term containing the $T^3 {\rm ln}(T)$
temperature dependence; analytic contributions also appear but again
these are not interesting.
The integral in the second line equals $\pi^4/30$ so the
final result is: ${{\delta C_V}\over{L^3}}
\approx -{{8}\over{15}}~ b~ {{k_B^4 m^4}\over{p_f^2}}
T^3 {\rm ln}(T)\ .$  Not surprisingly, this result has the same form as
that found by Pethick and Carneiro\refto{PC} as dimensional analysis
guarantees this.  A direct comparison of the coefficient is
meaningless however since our interaction is not the same.

\vfill\eject
\centerline{V. COLLECTIVE MODES}
\smallskip
The curvature of the Fermi surface did not play an important role in the
calculation of the specific heat.  In fact we
took the Fermi surface to
be flat; consequently the Hamiltonian could be rewritten as the
sum of products of a single creation and a single annihilation operator
(see Eq. [4.2]).  Collective
excitations of the Fermi surface, on the other hand, arise from the curvature.
It is therefore interesting to derive the spectrum of collective modes within
the new framework.  For a curved Fermi surface the Hamiltonian can
contain products, for example,
of two creation operators, so the more general Bogoliubov
transformation is required to diagonalize it.

Again we concentrate on the charged excitations to illustrate the quantum
theory.
We diagonalize the Hamiltonian Eq. [3.1] by first
taking the matrix square root of $V_c$
$$V_c({\bf S, T; q}) = \sum_{\bf U}~
V_c^\half({\bf S, U; q})~ V_c^\half({\bf U, T; q})~ ,\eqno(5.1)$$
then we rewrite the Hamiltonian as:
$$\eqalign{H &= \half~ \sum_{\bf q}~ \sum_{\bf U}~
[\sum_{\bf S}~ V_c^\half({\bf U, S; q})~ J({\bf S; q})]~
[\sum_{\bf T}~ V_c^\half({\bf U, T; q})~ J({\bf T; -q})] \cr
&= \half~
\sum_{\bf U}~ \sum_{\bf q}~ \tilde{J}({\bf U; q})~ \tilde{J}({\bf U; -q})
\ . \cr} \eqno(5.2)$$
Here we have introduced new charge currents:
$$\tilde{J}({\bf U; q}) \equiv \sum_{\bf S}~ V_c^\half({\bf U, S; q})~
J({\bf S; q})\eqno(5.3)$$
and also use the fact the $V_c$ is a real symmetric matrix [ie.
$V_c({\bf S, T; q}) = V_c({\bf T, S; q})$] so therefore $V_c^\half$ is
also symmetric.
These new currents obey modified Kac-Moody commutation relations:
$$[\tilde{J}({\bf S; q})~ ,~ \tilde{J}({\bf T; p})] =
2 \Omega~ \delta^3_{\bf p, q}
[V_c^\half D V_c^\half]({\bf S, T; q}) \eqno(5.4)$$
where the diagonal matrix
$$D({\bf S, T; q}) \equiv \delta^2_{\bf S, T}~
{\bf q \cdot \hat{n}_S} \eqno(5.5)$$
appears naturally in the implicit matrix product on the right hand side of the
equation.  (The sum over the internal indices in Eq. [5.4]
has been suppressed for
clarity.)  We obtain the spectrum by diagonalizing this modified anomaly.
Let the eigenvectors $u^{\bf A}({\bf S; q})$ and
eigenvalues $\omega^{\bf A}({\bf q})$ of the spectrum carry the label $\bf A$.
Suppressing again internal matrix indices and the momentum $\bf q$ we have:
$$2 \Omega~ \sum_{\bf T}~ [V_c^\half D V_c^\half]({\bf S, T})~
u^{\bf A}({\bf T})
= \omega^{\bf A} u^{\bf A}({\bf S})\ .  \eqno(5.6)$$
Upon matrix multiplying both sides of this equation by $V_c^\half$
and defining new eigenvectors $\tilde{u}^{\bf A}({\bf S}) \equiv \sum_{\bf T}
V_c^\half({\bf S, T})~ u^{\bf A}({\bf T})$ we arrive at the collective
mode equation (with an implicit sum over repeated indices):
$$2 \Omega~ V_c({\bf U, S; q}) D({\bf S, T; q}) \tilde{u}^{\bf A}({\bf T; q})
= \omega^{\bf A}({\bf q}) \tilde{u}^{\bf A}({\bf U; q})\ . \eqno(5.7)$$
This equation can be rephrased in a more familiar form by writing $V_c$
explicitely as $\half~ v_f~ \Omega^{-1}~ \delta^2_{\bf S, T}
+ {{1}\over{L^3}}~ V'_c({\bf S, T; q})$
and taking the interaction $V'_c$ to be independent of ${\bf q}$.
Dropping the label $\bf A$ and the tilde we find the
dispersion relation:
$$({\bf q \cdot v_S} - \omega)~ u({\bf S}) + {{2 \Omega}\over{L^3}}~
\sum_{\bf T}~
V'_c({\bf S, T})~ ({\bf q \cdot v_T})~ u({\bf T}) = 0\ .\eqno(5.8)$$
(Recall that ${\bf v_S} \equiv v_f {\bf \hat{n}_S}$ is the Fermi velocity
at grid point $\bf S$.)
Now we multiply each term in Eq. [5.8]
by $\bf q \cdot v_S$ and make another change of variable
by redefining $u({\bf T}) \rightarrow {\bf q \cdot v_T}~ u({\bf T})$
(with no sum over $\bf T$).  The result is:
$$({\bf q \cdot v_S} - \omega)~ u({\bf S})
+ {\bf q \cdot v_S}~ {{2 \Omega}\over{L^3}}~ \sum_{\bf T}~
V'_c({\bf S, T})~ u({\bf T}) = 0\ .\eqno(5.9)$$
Recognizing that the sum is just a coarse-grained version of the sum
over momenta $\bf k$ lying on the Fermi surface (FS):
$$\Omega~ \sum_{\bf T} \approx {{L}\over{2 \pi}}~ \sum_{\bf k \in FS}
\eqno(5.10)$$
we see that we have arrived at the collective mode equation.

Since zero sound excitations involve global distortions of the Fermi
surface that slosh Fermions back and forth, the curvature of the Fermi
surface plays an important role.  For example, solving this equation for
a perfectly spherical Fermi surface with $V'_c({\bf S, T})$ assumed to
be a constant independent of the angle between $\bf S$ and $\bf T$ we
find the zero-sound mode:
$$u(\theta, \phi) \propto {{{\rm cos}(\theta)}\over{s - cos(\theta)}}
\eqno(5.11)$$
where $(\theta,~ \phi)$ are polar coordinates with the polar axis in
the $\bf \hat{q}$ direction and $s \equiv {{\omega}\over{v_f |{\bf q}|}}$.
Also implicit in Eq. [5.9] is the renormalization of the Fermion mass.
Again assuming a spherical Fermi surface, we may use Galilean
invariance\refto{Baym} to find the well-known result:
$${{1}\over{m}} = {{1}\over{m^*}} + {{k_f}\over{3\pi^2}}~ f^c_1 \eqno(5.12)$$
if we identify $V'_c(\theta) = \sum_{l=0}^\infty~ f_l^c~ P_l({\rm cos}\theta)$.
Finally, collective excitations in the spin sector of our theory are given
by the corresponding Fermi liquid formula.
Apparently our new formulation of the Fermion liquid reproduces well known
Fermi liquid theory results.

\vfill\eject
\centerline{VI. DISCUSSION}
\smallskip

In the preceding sections we showed that a simple model of interacting Fermions
in two spatial dimensions can lead to a fixed point with local U(1)
symmetry despite the fact that the bare Hamiltonian is only invariant
under global U(1) transformations.
We also presented a framework for the Bosonization of Fermion
liquids in higher dimension.  We enlarge upon the connection between the
simplified model of section (II) and the general problem of Bosonization here.
First it is clear that the model
is pathological in the sense that the residual fixed point
current-current interactions that couple the four Fermi points are singular;
for example, $\lambda_{1c}$ (which couples currents at opposite points)
is typically of the same order as $v_c$ (which
couples currents at the same point).  In the
physical case of a continuous Fermi surface in spatial dimensions of two
or higher, interactions of this type would be equivalent to a current-current
coupling of the form:
$$V_c({\bf S, T; q}) = \half v_c \Omega^{-1}~ \delta^2_{\bf S,T}
+ \lambda_{1c}~ \Omega^{-1} \delta^2_{\bf S, -T}\eqno(6.1)$$
where $\bf -T$ denotes a mesh point directly opposite point $\bf T$.  In the
$\Lambda \rightarrow 0$ limit
of a finer and finer mesh, $\Omega^{-1} \propto \Lambda^{-2}
\rightarrow \infty$.  Thus the second term in Eq. [6.1] amounts to
a singular interaction that might be expected to destroy
Fermi-liquid type behavior\refto{Stamp}.

For singular interactions, however,
the connection between the multidimensional Bosonizaton and
one-dimensional behavior
begins to break down for at least two reasons.
First, as we noted in section (I), Luttinger liquids
in one spatial dimension
are characterized by the elimination of the discontinuity in the Fermion
occupancy at the Fermi surface.  Consequently, the Fermion distribution
is smeared out over some energy scale (set by the energy cut-off in the
interaction).  As long as this cut-off is small compared to lattice
energy scales, the continuum analysis of section (I) holds.  In higher
dimensions, however, a second energy scale $v_f \lambda$ has to be
introduced since
the Kac-Moody algebra is obtained in our construction only in the
$\omega$-limit of $\Lambda >> \lambda >> |{\bf q}|$ where $\Lambda
\rightarrow 0$.
This limit apparently precludes the
incorporation of interactions which eliminate the Fermi discontinuity.

We alluded to a second problem with singular spin-spin
interactions earlier: the
semi-classical limit breaks down because the terms on the right-hand
side of the spin current commutation relations Eq. [3.11] cannot be neglected
when interactions diverge in the $\Lambda \rightarrow 0$ limit.
A return to the original Luther Bosonization prescription using
narrow tubes instead of squat pill boxes appears to offer a way out of
both difficulties.  In this case the energy scale $v_f \lambda$ need not
be introduced; we can think of the higher-dimensional problem
as a collection of purely one-dimensional theories.  However,
now {\it only} singular ``tomographic\refto{PWA2}''
type interactions are permitted:
the current in any given tube can couple only to itself or to the current in
a tube
emerging from a point directly opposite on the Fermi surface.
Actually, the simplified model of section (II) illustrates this problem.
Interactions $\lambda_{2c}$ and $\lambda_{2s}$ only couple the zero-momentum
components of the currents and therefore cannot change the excitation spectrum.
Nevertheless the special case of tomographic Bosonization
may exhibit features of interest.

A separate, but related, problem of interest arises when
the velocity of charge excitations differs from that of spin excitations.
Fermi liquid theory breaks down in this case
because the Fermion propagator no longer
exhibits a simple pole; instead there is a branch cut.  Thus the quasi-particle
weight $Z = 0$ even though, as mentioned in section (I), a discontinuity in
the Fermion occupancy remains at the Fermi surface.
Since the two velocities are just
parameters appearing in $V_c({\bf S, T; q})$ and $V_s({\bf S, T; q})$ of our
theory, we need not restrict ourselves to setting both velocities
equal to a Fermi velocity $v_f$ as we did in section (III).
Anderson has argued
that a difference in velocities between the two sectors, rather than
singular interactions at the antipode, might account for the anomolous
normal state properties of the copper oxide superconductors\refto{PWA2}.
It might be interesting to explore the consequences of spin-charge separation
within the Bosonization framework presented here.
Other open problems include the incorporation of van Hove
singularities and energy gaps within the Bosonization framework.  It may also
be possible to include fluctuations in the Fermi surface shape or topology
within a renormalization group approach.

Finally, the Bosonization procedure outlined in this paper
may permit the application of semiclassical approximations to the
interacting Fermion problem.  Semiclassical approximations cannot be directly
applied to Fermions because the Pauli exclusion principle guarantees
that occupation numbers are of order one and thus far from the classical limit.
Bosonization bypasses this problem by replacing the Fermion variables
with Bosonic ones.  Indeed, semiclassical approximations to Bosonized versions
of certain one-dimensional problems have been remarkably useful in
the past. For example, an analysis of quantum spin chains that begins with
the weak-coupling Hubbard model describing interacting electrons ends up
mapping the low-energy theory onto the non-linear WZW sigma model.  The
behavior of this model in the semi-classical limit explains many of the
known properties of quantum antiferromagnets\refto{AH}.

\bigskip
\centerline{Acknowledgements}
We thank Ian Affleck, Sasha Finkel'stein, Matthew Fisher, Duncan Haldane,
Mike Kosterlitz, Hyok-Jon Kwon,
Patrick Lee, Andreas Ludwig, M. Randeria, N. Read, and R. Shankar
for helpful discussions.
J.B.M. thanks the Institute
for Theoretical Physics (NSF Grant No. PHY89-04035)
for its hospitality during his stay there.
A. H. was supported by the National Science
Foundation through grant DMR9008239.

\smallskip

\vfill\eject


\references

\refis{GO}Peter Goddard and David Olive, {\it Int. J. of Mod. Phys. A}
	{\bf 1}, 303 (1986).

\refis{LW}E. Lieb and F. Wu, {\it Phys. Rev. Lett.} {\bf 20}, 1445 (1968).

\refis{Emery}For a review, see V. J. Emery in {\it Highly Conducting
	One-dimensional Solids}, eds. J. T. Devrese, R. P. Evrard,
	and V. E. Van Doren (Plenum 1979).

\refis{Photo}J. C. Campuzano {\it et al.}, {\it Phys. Rev. Lett.} {\bf 64},
	2308 (1990).

\refis{Haldane0}F. D. M. Haldane, seminar at Brown University,
	November 1991.

\refis{Haldane2}F. D. M. Haldane, {\it J. Phys. C} {\bf 14}, 2585
	(1981).

\refis{Luther}A. Luther, {\it Phys. Rev. B} {\bf 19}, 320 (1979).

\refis{Shankar1}R. Shankar, {\it Physica A} {\bf 177}, 530 (1991).

\refis{AM}Ian Affleck and J. Brad Marston, {\it J. Phys. C} {\bf 21},
	2511 (1988).

\refis{Shankar2}For a good review of Abelian Bosonization,
	see R. Shankar, ``Bosonization:  how to make it work
	for you in condensed matter,'' lectures given at the BCSPIN
	school in Katmandu, Nepal, May 1991.

\refis{Shankar3}R. Shankar, private communication.

\refis{PWA}P. W. Anderson, {\it Science} {\bf 235}, 1196 (1987).

\refis{Ludwig}E. Witten, {\it Commun. Math. Phys.} {\bf 92}, 455 (1984);
	A. B. Zamolodchikov and V. A. Fateev, {\it Sov. J. Nucl. Phys.}
	{\bf 43}, 657 (1986); D. Gepner and E. Witten, {\it Nucl. Phys.}
	{\bf B278}, 493 (1986);
	A. W. W. Ludwig and Ian Affleck, {\it Nucl. Phys. B}. {\bf 360},
	641 (1991).

\refis{Ian}We thank I. Affleck for pointing this fact out to us.

\refis{PC}C. J. Pethick and G. M. Carneiro, {\it Physical Review A}
	{\bf 7}, 304 (1973).

\refis{Baym}G. Baym and C. Pethick, {\it Landau Fermi-Liquid theory:
	concepts and applications} (John
	Wiley \& Sons, New York, 1991).

\refis{Greywall}Dennis S. Greywall, {\it Phys. Rev. B} {\bf 27},
	2747 (1983).

\refis{Schulz}H. J. Schulz, {\it Europhysics Lett.} {\bf 4}, 609 (1987).

\refis{Dzy}I. E. Dzyaloshinskii, {\it Sov. Phys. JETP} {\bf 66}, 848 (1987);
	I. E. Dzyaloshinskii and V. M. Yakovenko, {\it Sov. Phys. JETP}
	{\bf 67}, 844 (1988).

\refis{PWA1}P. W. Anderson, {\it Phys. Rev. Lett.} {\bf 67}, 3844 (1991).

\refis{Fink}A. Finkel'stein, unpublished.

\refis{Cast}C. Castellani, C. Di Castro, and W. Metzner, {\it Phys. Rev. Lett}
	{\bf 69}, 1703 (1992).

\refis{Stamp}P. C. E. Stamp, {\it Phys. Rev. Lett.} {\bf 68}, 2180 (1992).

\refis{PWA2}P. W. Anderson, {\it Phys. Rev. Lett.} {\bf 67}, 2092 (1991).

\refis{AH}Ian Affleck and F. D. M. Haldane, {\it Phys. Rev. B} {\bf 36},
	5291 (1987).

\endreferences

\vfill\eject


\head{Figure Captions}

\item{(1)} Typical band structure and the left and right Fermi points of the
	one-dimensional problem.  The dashed lines denote filled states.

\item{(2)} A one-loop diagram that gives a second order
	contribution to the renormalization of $\lambda_s$.
	All diagrams that contribute at this order have
	a left and a right moving propagator which
	connect the point (x, t) with the point (0, 0).

\item{(3)} The four Fermi points kept in the model with linear
	dispersion along the lines.  The dotted line delineates the
	Fermi surface of the half-filled nearest-neighbor tight
	binding model.  The inset shows the (u, v) coordinate system.

\item{(4)} The nine types of marginal interactions:
	(a) The four current-current interactions that
	respect spin-charge separation.
	(b) The one non-Umklapp mixed process.
 	(c) The two Umklapp processes that transport charge-2 spin
	singlets and therefore respect spin-charge separation.
	(d) The two other ``mixed'' Umklapp processes that
	transport both spin and charge and thus break spin-charge separation.
	(The Umklapp processes only occur at half-filling.)
	Note that these diagrams only depict representative
	process -- the missing diagrams are generated by performing the
	various symmetry operations on the square lattice
	(reflections that exchange points 1 and -1 or 2 and -2
	and rotations through 90 degrees).

\item{(5)} Renormalization group flow of the model away from
	half-filling.  The initial coupling is a repulsive Hubbard
	interaction with $U/t = 1$.  The couplings flow toward the fixed line
	$\lambda_{1s} = \lambda_5 = 0$ and $\lambda_{1c} > 0$.

\item{(6)} Currents at each grid point ${\bf S}$ on the Fermi surface are
	constructed with the use of squat pill boxes that tile the surface.
	The box has dimensions $\Lambda \times \Lambda$ along the surface,
	height $\lambda = {{\Lambda}\over{\sqrt{N}}}$
	(where $N \rightarrow \infty$)
	and is bisected through the mid-plane by the Fermi Surface.
	The function $\theta({\bf S; k}) = 1$ inside the box; otherwise
	it is zero.
	The momentum $\bf q$ must be small: $|{\bf q}| \leq
	{{\Lambda}\over{N}}$.  Thus, $|{\bf q}| << \lambda << \Lambda$.

\item{(7)} Geometry of the Fermi liquid interactions that lead to
	non-analytic contributions to the specific heat.  Two
	squat boxes lie on the locally flat Fermi surface at grid points
	${\bf S}$ and ${\bf T}$ (see text).

\vfill\eject
\end